\documentclass[a4paper]{ar-1col}

\usepackage{graphicx}
\usepackage{cancel}
\usepackage{natbib}
\usepackage{amsmath,amsbsy}

\jname{Ann. Rev. Astron. Astrophys.}
\jvol{ARAA}
\jyear{2016}

\def\be{\begin{equation}}
\def\ee{\end{equation}}
\def\lesssim{\mathrel{\rlap{\lower4pt\hbox{\hskip1pt$\sim$}}
    \raise1pt\hbox{$<$}}}         
\def\gtrsim{\mathrel{\rlap{\lower4pt\hbox{\hskip1pt$\sim$}}
    \raise1pt\hbox{$>$}}}         
\def\VEV#1{{\left\langle #1 \right\rangle}}

\newcommand{\vsp}{\vphantom{\Big[}\\}
\newcommand{\Mpl}{{M_{\rm Pl}}}
\newcommand{\calR}{{\cal R}}

\begin{document}

\markboth{Kamionkowski \& Kovetz}{Cosmic B modes and Inflation}

\title{The Quest for B Modes from Inflationary Gravitational Waves}

\author{Marc Kamionkowski and Ely D. Kovetz
\affil{Department of Physics and Astronomy, Johns Hopkins
University, 3400 N.\ Charles St., Baltimore, Maryland, 21218, USA}}

\begin{abstract}
The search for the curl component (B mode) in the cosmic microwave background (CMB) polarization induced by inflationary gravitational waves is described.  The canonical single-field slow-roll model of inflation is presented, and we explain the quantum production of primordial density perturbations and gravitational waves.  It is shown how these gravitational waves then give rise to polarization in the CMB.  We then describe the geometric decomposition of the CMB polarization pattern into a curl-free component (E mode) and curl component (B mode) and show explicitly that gravitational waves induce B modes.  We discuss the B modes induced by gravitational lensing and by Galactic foregrounds and show how both are distinguished from those induced by inflationary gravitational waves.  Issues involved in the experimental pursuit of these B modes are described, and we summarize some of the strategies being pursued. We close with a brief discussion of some other avenues toward detecting/characterizing the inflationary gravitational-wave background.
\end{abstract}

\maketitle

\tableofcontents

\section{Introduction}

The daily vistas encountered by the inhabitants of Earth feature
seashores, mountains, cliffs, ice, fire, raging storms, and a
sky that meets the ground through a horizon that may be jagged
or straight.  Our astronomical vista is similarly complex, with
a startling display of stars, binary stars, compact objects,
intergalactic gases and dust, an array of galaxies, and clusters
of galaxies, no two of which look precisely alike. Our cosmic
vista, however, is far simpler.  When we look to the greatest
observable cosmic distances, the Universe is virtually the same
everywhere, with only tiny departures from homogeneity.  It
turns out that these subtle inhomogeneities, which through
heroic experimental efforts have now been mapped with
formidable precision, exhibit nontrivial but
still surprisingly simple patterns. According to the standard 
cosmological model, these are seeded by primordial quantum 
perturbations that are imprinted onto the early Universe by a 
process of rapid exponential expansion known as inflation.

The era of precision cosmology was ushered in during the 
past decade by ever more accurate measurements of the 
distribution of mass in the Universe on cosmic distance scales 
using galaxy surveys \citep{Dawson:2012va, Amendola:2012ys} and through 
measurements of temperature fluctuations and polarization in 
the cosmic microwave background (CMB)
\citep{Bennett:2012zja,Adam:2015rua}. 
Together, these measurements
indicate that the early Universe was homogeneous to better than
one part in 10,000.  They indicate that departures from
homogeneity are ``adiabatic''---i.e., they preserve the ratios
of the different components of matter (baryons, dark matter,
radiation, and neutrinos) in the Universe.  As discussed below,
the primordial inhomogeneities (which hereafter we refer to as
density perturbations) are well described by an impressively
simple structure, a nearly scale-invariant
spectrum with a Gaussian distribution of Fourier amplitudes.

The goal of early-Universe cosmology is to quantify the
observed features of the Universe and to develop a physical
model to account for them.  Up until several decades ago, the
number of such models was huge, and they made a
large array of predictions for the nature of primordial
perturbations: for the Fourier-space spectrum; for correlations
between the different Fourier amplitudes; for spatial structures
like cosmic strings, textures, and monopoles; for fluctuations
in the ratios of baryons, dark matter, radiation, and/or
neutrinos, etc.  Today, the vast majority of these models are dead,
ruled out by the onslaught of precise measurements.  The overwhelming
majority of those that survive involve inflation
\citep{Brout:1977ix,Starobinsky:1980te,Kazanas:1980tx,Sato:1980yn,Guth:1980zm,Linde:1981mu,Albrecht:1982wi}, a
period of accelerated expansion in the very early
Universe (within the first fraction of a nanosecond of the
Universe), and explain primordial perturbations as quantum
fluctuations in the spacetime metric during inflation
\citep{Mukhanov:1981xt,Hawking:1982cz,Guth:1982ec,Starobinsky:1982ee,Linde:1982uu,Bardeen:1983qw}.

The simplest and canonical model for inflation---namely
single-field slow-roll (SFSR) inflation---made a number of
predictions that have been confirmed by a sequence of
increasingly precise experiments over the past two decades.
These include the prediction that (a) primordial perturbations
are adiabatic; (b) the spectrum of primordial
perturbations should be very nearly scale invariant, but not
precisely scale invariant; (c) the distribution of primordial
perturbations should be very nearly Gaussian; and (d) there
should be primordial perturbations that are super-horizon (i.\ e.\ with
wavelengths larger than the Hubble radius) at the
time of CMB decoupling.  The consistency of these predictions
with all current cosmological data suggests that inflation is an
idea that should be taken seriously and studied further.

Still, inflation raises its own set of questions (e.g., what set
it in motion?), and the literature is teeming with detailed
implementations.  The focus of early-Universe cosmology in the
forthcoming years will be to further test the notion of inflation
and narrow the range of inflationary models and scenarios.
Given that we are talking about physics from 13.8
billion years ago, when the relevant energy scales were well beyond
those at accelerator laboratories, this is an ambitious quest.
Since observable fossils from that time are few and far between,
any conceivable empirical avenue to inflation should be
pursued.

In addition to the predictions for primordial density perturbations
discussed above, SFSR inflation also predicts the existence of a
stochastic background of gravitational waves
\citep{Starobinsky:1979ty,Rubakov:1982df,Fabbri:1983us,Abbott:1984fp}
that then induce a specific gradient-free ``B-mode'' pattern 
in the polarization of the CMB
\citep{Kamionkowski:1996zd,Seljak:1996gy,Kamionkowski:1996ks,Zaldarriaga:1996xe,Seljak:1996ti}.
This review focuses on these B modes and their role in
addressing the physics of inflation.  When first considered in
1996, the amplitude of these B modes could have been just about
anything, in the best-case scenario easily detectable and in the
worst-case scenario way too small to ever be seen.  As we
shall discuss, though, recent measurements of the spectral index
\citep{Knox:1995dq, Jungman:1995bz, Komatsu:2008hk,Calabrese:2013jyk,Ade:2013zuv,Ade:2015xua}
suggest, within the context of SFSR inflation, that the B-mode
signal may be strong enough to be detectable by experiments
planned for the next 5--10 years, making this a particularly
exciting time.

Below we begin by reviewing the basics of inflation, starting
with the homogeneous cosmic evolution during inflation and then
discussing the generation of density perturbations and gravitational waves.  We
discuss how gravitational waves produce polarization in the
CMB, and then describe the decomposition of the polarization
pattern into two distinct geometric components, a curl-free (E
mode) part and a curl (B mode) part.  We do so first within the
context of a flat-sky approximation before moving to the
full-sky formalism.  Next, we show explicitly that gravitational
waves give rise to B modes while primordial density
perturbations do not (at linear order in the perturbation
amplitude).  We then discuss the B modes induced by lensing of
the CMB by density inhomogeneities along the line of sight
\citep{Zaldarriaga:1998ar} and
also how these lensing-induced B modes can be distinguished from
primordial B modes \citep{Kesden:2002ku,Knox:2002pe}.  The
remainder of the article discusses the
detectability of the signal, strategies for detection, and
issues involved in separating a cosmic signal from that due to
Galactic foregrounds.  
We close with a brief discussion of some other avenues 
toward detection and characterization of the inflationary 
gravitational-wave background.

Before proceeding with our review, we provide an incomplete list
of reviews of related subjects.  An early review
\citep{Brandenberger:1984cz} and much of the formalism for
cosmological perturbations was elaborated in
\citep{Kodama:1985bj,Mukhanov:1990me,Bertschinger:1993xt} and
then updated in \citep{Malik:2008im}.  \citet{Olive:1989nu}
reviewed inflation models around 1990 followed by
\citet{Lyth:1998xn} in 1997.  
\citet{Lidsey:1995np} reviewed single-field slow-roll inflation
and then several more recent articles review models for
inflation beyond single-field slow roll.  These include a review
of curvaton models \citep{Mazumdar:2010sa}, one about models with
gauge fields \citep{Maleknejad:2012fw}, and others
of models that embed inflation in string theory
\citep{Baumann:2009ni, Baumann:2014nda, Westphal:2014ana}.   
\citet{Martin:2013tda} classifies a
broad range of inflationary models.  \citet{Bartolo:2004if}
discusses non-gaussianity and inflation.  The recent Planck
inflation papers \citep{Planck:2013jfk,Ade:2015lrj} also provide
very nice up-to-date discussions of inflation.
\citet{Kamionkowski:1999qc} reviewed connections between particle
physics and the CMB; \citet{Hu:2001bc} reviewed the theory of CMB
fluctuations; and \citet{Lewis:2006fu} discussed lensing of the
CMB.  \citet{Hu:1997hv} provided a short but elegant early
review of CMB polarization, and this review builds and expands
upon an earlier review \citep{Cabella:2004mk}.  The review of
\citet{Samtleben:2007zz} on CMB polarization complements this
review in its deeper coverage of experimental techniques.
Finally, \citet{Buonanno:2014aza} provides a recent review of
the direct search for gravitational waves.

\section{Inflation basics}
\label{sec:inflationbasics}
Inflation has become such a dominant paradigm that we often
forget the original motivations---the flatness problem (why is
the present ratio of the energy density relative to 
the critical energy density so close to unity?), the horizon problem (why do
causally disconnected regions at the CMB surface of last scatter
have the same temperature), and the monopole problem
\citep{Preskill:1979zi}---at the time, $\sim1980$, that the idea
began to take shape.  The solution to all these problems was
provided by a postulated period of accelerated expansion in the
early Universe \citep{Guth:1980zm}.

\subsection{Homogeneous evolution}

\subsubsection{Kinematics}
An expanding isotropic and homogeneous Universe is described by
a Friedmann-Robertson-Walker (FRW) spacetime, with line element
$ds^2=-dt^2 + a^2(t) d\boldsymbol x^2$, in terms of a scale factor
$a(t)$ that parametrizes the physical distance that corresponds
to a given comoving distance.
As the Universe expands [i.e., the scale factor $a(t)$
increases with time $t$], the Hubble length $H^{-1}$, where
$H\equiv \dot a/a$ is the Hubble or growth rate, increases.
During radiation and matter domination, $(d/dt)(aH)^{-1} > 0$,
and so the Hubble distance $H^{-1}$ increases more rapidly
than the scale factor.  As a result, with time, an observer
sees larger comoving volumes of the Universe, and objects and
information enter the horizon.  This observation leads to
the horizon problem: if the Universe began with a period of
radiation domination, then how did the $\sim40,000$ causally
disconnected patches of CMB sky know 
to have the same temperature (to roughly one part in 10,000)?

If, however, $(d/dt)(aH)^{-1}  <0$, then an observer sees with time a smaller comoving
patch (even though the physical or proper size of the observable
patch may still be increasing), and
objects/information/perturbations exit the horizon.  In this
way, the Universe becomes increasingly smooth, thus explaining
the remarkable large-scale homogeneity of the Universe. 

The requirement $(d/dt)(aH)^{-1} = \left[(\dot H/H^2)+1
\right]/a<0$, where the dot denotes a derivative with respect to 
time, implies that we must have $\epsilon \equiv -\dot H/H^2 <1$
for inflation.  Most generally, $\dot H \neq0$ (so that
inflation can end, if for no other reason).  As we will see,
however, theory and measurement suggest $\epsilon \ll1$,
implying that the scale factor grows almost exponentially, $a(t)
\propto e^{Ht}$, during inflation. 

If we assume the validity of general relativity, as we do here
[although there is a vast literature on inflation with
alternative gravity theories; e.g.,
\citet{La:1989za,DeFelice:2011zh,Clifton:2011jh}],
then the time evolution of the
scale factor satisfies the Friedmann equations, $H^2=\rho/(3
\Mpl^2)$ and $\dot H+H^2 = -(\rho+3p)/(6 \Mpl^2)$,
where $p$ and $\rho$ are the pressure and energy density of the
cosmic fluid, respectively.  We work in particle-physics units,
with $\hbar=c=1$ and have written Newton's constant $G=(8\pi
\Mpl^2)^{-1}$ in terms of the reduced Planck mass,
$\Mpl=2.435\times 10^{18}$~GeV.  These two Friedmann equations
imply that
\begin{equation}
    \epsilon = (3/2) \left( 1 + p/\rho\right),
\end{equation}
from which we infer that the equation-of-state parameter
$w\equiv p/\rho$ must be $w<-1/3$ in order for inflation to
occur. 

\subsubsection{Scalar-Field Dynamics}
\begin{figure}
\includegraphics[width=\linewidth]{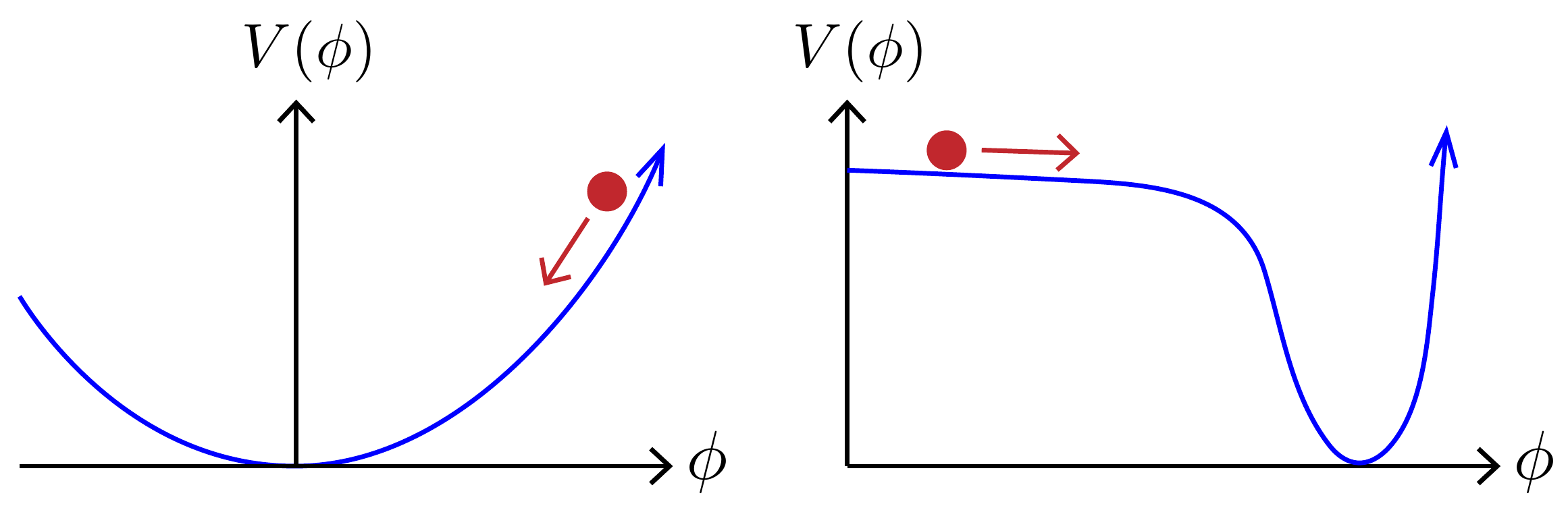}
\caption{Inflation postulates that at some point in the early history of the Universe, the cosmic energy density was dominated by the vacuum energy associated with the displacement of some scalar field $\phi$ (the ÒinflatonÓ) from the minimum of its potential.  Shown here for illustration are two toy models for the inflaton potential: on the left, a quadratic potential and on the right, a ÒhilltopÓ potential.  (Adapted from \protect\citet{Kamionkowski:1999qc}).}
\label{fig:potentials}
\end{figure}

In the simplest paradigm for inflation, and that on
which we focus, this exotic equation of state is provided by the
displacement of a scalar field $\phi$, the ``inflaton,'' from
the minimum of its potential $V(\phi)$.  The homogeneous
time evolution of the scalar field then satisfies, in an FRW
spacetime, the equation of motion, $\ddot \phi + 3 H \dot\phi +
V'(\phi)=0$, where the prime denotes a 
derivative with respect to $\phi$.  We thus see that the
expansion acts as a friction term.  The scalar field has energy
density $\rho=(1/2) \dot\phi^2 +V(\phi)$ (a kinetic-energy
density and a potential-energy density) and pressure $p=(1/2)
\dot\phi^2 -V(\phi)$.  If $V(\phi)$ is nonzero and
sufficiently flat and the friction term in the $\phi$ equation
sufficiently large, then the kinetic-energy density will be
 $(1/2)\dot\phi^2 < 2 V(\phi)$, in which case
$p<-\rho/3$ and inflation ensues (see {\bf Figure \ref{fig:potentials}}).

This condition is made more precise by solving the scalar-field
equation of motion along with the Friedmann equation, $H^2 =
\left(\dot a/a\right)^2= \left[ V(\phi) + (1/2) \dot \phi^2
\right]/( 3m_{\rm Pl}^2)$.
During inflation $\phi$ varies monotonically with time $t$ and
can thus be used as the independent variable (rather than $t$).
Let us suppose that the field and potential are defined so
that $\dot\phi>0$ during inflation.  We then differentiate the
Friedmann equation with respect to time, obtaining 
$2 H \dot H = \dot \phi \left [V'(\phi) + \ddot \phi \right]/(2 m_{\rm Pl}^2)$.
Then rearranging the scalar-field equation of motion, $-3 H \dot
\phi = \ddot \phi + V'(\phi)$, we get $\dot H = \dot
\phi^2/(2 m_{\rm Pl}^2)$.  We thus infer that
\begin{equation}
     \epsilon = 3 \frac{ \dot \phi^2 /2}{ V+\dot\phi^2/2}\simeq
\frac{\Mpl^2}{2}\left(\frac{V'}{V}\right)^2,
\end{equation}
where the last expression is the result of the slow-roll approximation, 
$\epsilon \ll1$, in which $\dot\phi^2/2 \ll V$. Note that in
much of the literature,
$\epsilon$ is ${\it defined}$ in terms of $V$ and $V'$ through
this relation, rather than through $\epsilon = -\dot H/H^2$, as
is done here, a distinction whose subtlety will be unimportant
in this article, although it can be important for quantitative
conclusions given the precision of current measurements.
We also define a second slow-roll parameter,
\begin{equation}
     \eta = -2 \frac{ \dot H}{H^2}- \frac{\dot
     \epsilon}{2H\epsilon} \simeq \Mpl^2 \frac{V''}{V},
     \label{eq:eta}
\end{equation}
which will become important below; the
approximation in Equation \ref{eq:eta} is valid during slow-roll, when $\eta \ll 1$.

\subsubsection{Duration of inflation and evolution of scales}
The number of $e$-folds of inflation between the end of
inflation and a time $t$ during inflation is
\begin{equation}
     N(t) \equiv \ln \frac{a(t_{\rm end})}{ a(t)} = \int_t^{t_{\rm
     end}}\, H \, dt = -\frac{1}{2 m_{\rm Pl}^2}
     \int_{\phi_t}^{\phi_{\rm end}} \, \frac{H}{ H'}\, d\phi
     = \int_{\phi_{\rm end}}^{\phi_t}\,
     \frac{d\phi}{\Mpl}\frac{1}{\sqrt{2\epsilon(\phi)}}.
     \label{eq:amount}
\end{equation}
The largest comoving scales exit the
horizon first during inflation, and they are the last to
re-enter the horizon later during matter or radiation domination.
To evaluate the number of $e$-foldings required to
solve the horizon problem, consider a physical wavenumber
$k_{\rm phys}$.  Its ratio to the Hubble scale today is
\begin{equation}
     \frac{ k_{\rm phys}}{ a_0 H_0} = \frac{a_k H_k}{ a_0 H_0} =
     \frac{a_k}{ a_{\rm end}} \frac{a_{\rm end}}{ a_{\rm reh}} \frac{
     a_{\rm reh}}{ a_{\rm eq}} \frac{a_{\rm eq}}{ a_0} \frac{H_k}
     { H_0},
\end{equation}
where $a_k$ and $H_k$ are the scale factor and Hubble parameter
when this particular wavenumber exits the horizon; $a_{\rm end}$
is the scale factor at the end of inflation; $a_{\rm eq}$
is the scale factor at matter-radiation equality; and $a_{\rm eh}$ is
the scale factor at the time of reheating.  Plugging in
numbers, we find that the number of $e$-foldings between the end
of inflation and the time at which the wavenumber $k$ exits the
horizon is
\begin{equation}
     N(k) = 62 - \ln \frac{k_{\rm phys}}{ a_0 H_0} - \ln
     \frac{10^{16}\, {\rm GeV}}{ V_k^{1/4}} + \ln\frac{ V_k^{1/4}}
     { V_{\rm end}} -\frac13 \ln \frac{V_{\rm end}^{1/4}}
     {\rho_{\rm reh}^{1/4}},
\end{equation}
where $\rho_{\rm reh}^{1/4}$ is the energy density at reheating. 
If the energy scale of inflation is near the current upper limit
$V^{1/4} \lesssim 10^{16}$~GeV (see below), but higher than the
energy scale of electroweak symmetry breaking ($V_k\gtrsim
10^{3}$~GeV), then the number $N$ of $e$-folds between the time that the
largest observable scales today exited the horizon and the end of inflation falls in the range
$30 \lesssim N\lesssim 60$.  Recent treatments that consider
different families of inflationary potentials, include
current constraints to the scalar spectral index $n_s$ (see
below), as well as plausible reheating scenarios, find a
range $40\lesssim N \lesssim 60$
\citep{Dai:2014jja,Cook:2015vqa,Munoz:2014eqa}.  More
conservatively, the near
scale-invariance of primordial density perturbations over the
$\sim3$ orders of magnitude over which they have been measured
tells us that $N\gtrsim10$ at the very least.

\begin{figure}
\includegraphics[width=1\linewidth]{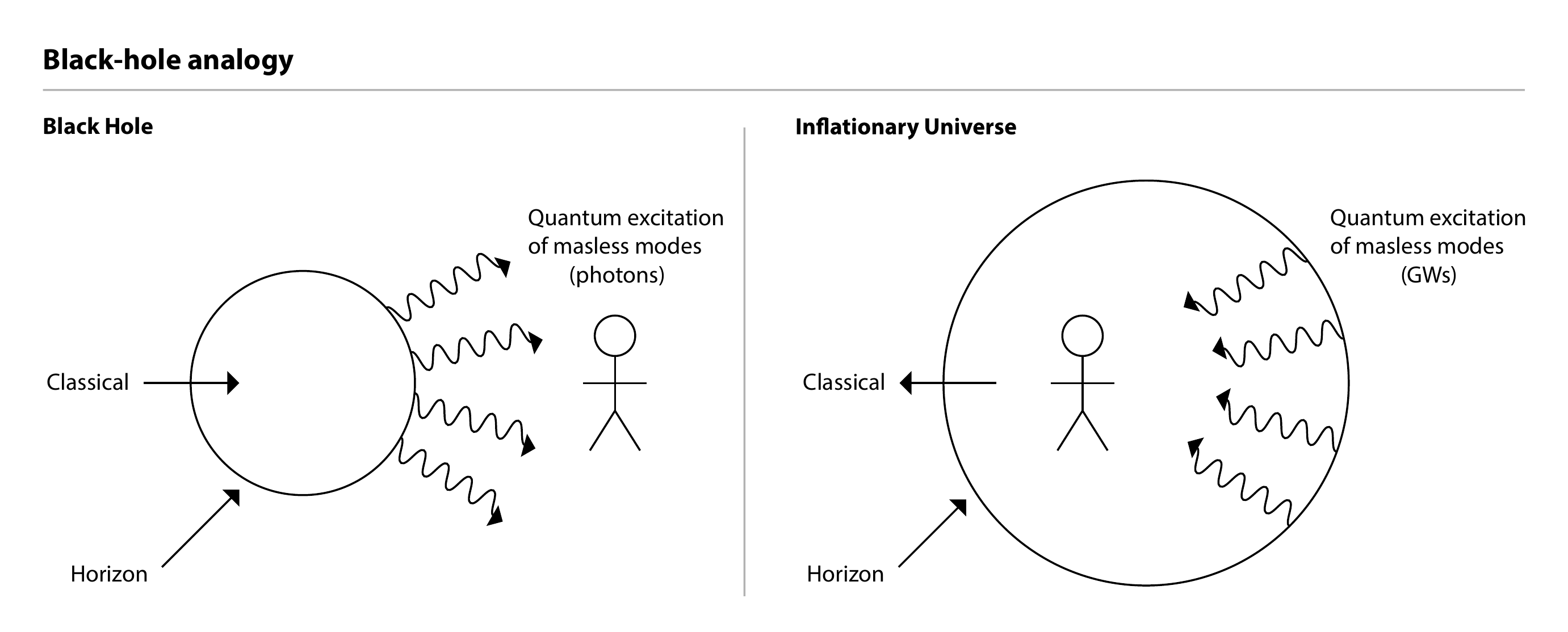}
\caption{{\it Left}: Hawking radiation from a black-hole event horizon;
     {\it Right}: Hawking radiation from an event horizon in an
     accelerating FRW spacetime. The ``classical" arrows indicate the direction of information flow according to classical general relativity.}
\label{fig:heuristic}
\end{figure}

\begin{textbox}
\subsubsection{Heuristic understanding of inflationary gravitational
waves}
Here we present in simple heuristic terms the
origin of inflationary gravitational waves (IGWs).
Consider first a black hole.  As shown in {\bf Figure \ref{fig:heuristic}}, 
it has an event horizon, a spherical surface
beyond which, according to (classical) general relativity, objects
and information disappear without a trace.  Hawking showed, 
however, that when quantum mechanics is
taken into consideration, the horizon glows---it emits
electromagnetic radiation \citep{Hawking:1974sw}.  Hawking's
argument also applies, however, to any radiation field with
massless quanta, and so the black hole also radiates
gravitational waves. In an FRW Universe with an accelerated expansion, 
there is also a horizon, a spherical surface beyond which 
(according to general relativity) objects and 
information disappear.  This time, though, the observer sees
this spherical surface from the inside, rather than outside.
Just as was the case with the black-hole horizon, this horizon
also radiates gravitational waves, according to quantum
mechanics.  These gravitational waves are produced
throughout inflation, and the expansion rate and thus
horizon temperature are nearly constant during inflation.  These
gravitational waves thus remain, after inflation, as a
primordial-gravitational-wave background with a nearly 
scale-invariant spectrum.

\end{textbox}

\subsection{Density (scalar metric) perturbations}
\label{sec:densityperts}
We now discuss the production of primordial density
perturbations from quantum fluctuations in the inflaton.
Although this involves a straightforward application
of the techniques of quantum fields in curved spacetime
\citep{Birrell:1982ix}, the precise calculation involves a level
of technical detail beyond the scope of this work.  Here we
therefore only outline the calculation schematially and refer
the reader to one of the many good pedagogical references [e.g.,
\citet{Lyth:1998xn,Liddle:2000cg,
Dodelson:2003ft,Mukhanov:2005sc,Weinberg:2008zzc,Baumann:2014nda}]
for the technical details.

If the scalar field $\phi$ or Hubble parameter (which is
determined at any point by the scalar-field value at that point)
can vary in time, then they can also vary in space.  We thus
consider the perturbations to the spacetime metric induced by
spatial fluctuations in the scalar field.  Since the energy
density is determined by $\phi$, fluctuations in $\phi$ will
induce fluctuations in the energy density which will then induce
fluctuations in the spacetime metric.  Although the most general
metric perturbation has ten components, four are unphysical
gauge modes.  Of the remaining six, two are tensor degrees of
freedom and two are vector degrees of freedom, none of which can
be sourced by perturbations to the scalar field.  It can further
be shown \citep[e.g.,][]{Mukhanov:1990me} that the remaining two
scalar degrees of freedom are reduced to one for scalar-field
perturbations.  In the comoving gauge, $g_{0\mu}=0$, the spatial
components of the metric in a scalar-field-dominated Universe
are written, $g_{ij} = a^2(t) {\rm exp} \left[2 \calR(\boldsymbol x,t)
\right] \delta_{ij}$, 
in terms of the ``curvature perturbation'' $\calR(\boldsymbol x,t)$.
Inserting this metric into the Einstein-Hilbert action, combined
with the action for the scalar field, and expanding to quadratic 
order in the perturbation $\calR$, we get
\begin{equation}
     S_\calR = \int\, dt\, \int\, d^3\boldsymbol x \, a^3  \left[
     (1/2)\dot v^2 - (1/2)(\nabla v)^2/a^2 \right],
\label{eqn:calRaction}
\end{equation}
in terms of a new field variable $v^2 = 2\Mpl^2 \epsilon \calR^2$. 
Here $\boldsymbol x$ is a comoving coordinate and $\nabla$ is a
gradient with respect to $\boldsymbol x$.  
We now Fourier transform the spatial part of $v$ to
write,
\begin{equation}
     S_\calR = \sum_{\boldsymbol k} \int\, dt \, a^3 \left[(1/2) |\dot
     v_{\boldsymbol k}|^2 - (1/2)(k/a)^2 |v_{\boldsymbol k}|^2 \right],
\label{eqn:oscillatoractions}
\end{equation}
which we recognize as the sum of actions for an ensemble 
of uncoupled oscillators, one for each $\boldsymbol k$, and each with
frequency $k/a$.  

This can be seen, if we begin with a Lagrangian, $L=(1/2) \dot
v^2-(1/2) k^2 v^2$, by making the 
variable substitutions $v=\sqrt{m} x$ and $k^2 = \omega^2=
\kappa/m$.  We then obtain the Lagrangian $L=(1/2)m \dot x^2 -
\kappa x^2$ for a simple harmonic oscillator of displacement $x$, mass
$m$, spring constant $\kappa$, and angular frequency $\omega$.
We also know that in both the quantum and classical treatments,
the average kinetic and potential energies are equal, and equal
to half the total energy.  Moreover, in the quantum-mechanical 
ground state, these are both found to be $\hbar
\omega/2$.  It thus follows that the probability density to find 
the oscillator with amplitude $v$ is Gaussian,
with variance $\VEV{v^2} = \hbar/(2\omega)$.

Now return to Equation \ref{eqn:oscillatoractions}.  Variation of
the action for each Fourier mode $\boldsymbol k$ results in an equation
of motion,
\begin{equation}
     \ddot v_{\boldsymbol k} +  3 H \dot v_{\boldsymbol k} + (k/a)^2
     v_{\boldsymbol k}=0,
\label{eqn:veom}
\end{equation}
for the time evolution of each $v_{\boldsymbol k}$.  This is the
equation of motion for a simple harmonic oscillator with a
time-dependent frequency $k/a$ (the time dependence arises
because of the stretching of the wavelength of a comoving
Fourier mode) and a friction term $3H$ that arises from the
expansion.  Consider the
system at some range of times $t$ centered around a time $T$
well before horizon exit ($k/a \gg H$) with a spread of times
$|t-T| \ll H^{-1}$.  Over this range of times, the cosmic expansion
is negligible, as is the friction, and the
classical solution is simply sinusoidal.  The amplitude of the
oscillation, in the ground state, is fixed by the requirement
$\VEV{v^2} = 1/[2(k/a)]$ expected from our discussion
of the quantum simple harmonic oscillator.  We thus identify
$v_{\boldsymbol k}(t) = [2 E_k]^{-1/2} e^{-i E_k }$, for $k\gg aH$,
as the early-time mode function for the ground state of the
oscillator, where we have identified $E_k=k/a$ as the energy and
$\hbar=1$.  The complete solution to Equation \ref{eqn:veom}, with
this normalization at early times, is
\begin{equation}
     v_{\boldsymbol k}(t) = \frac{H}{(2k^3)^{1/2}} \left(i
     +\frac{k}{aH} \right) e^{-ik/aH}.
\end{equation}
We see that at late times, after horizon crossing ($k\ll aH$), the
solution approaches $|v_{\boldsymbol k}|^2 \to H^2/(2k^3)$.  From this
we infer that the inflationary expansion converts subhorizon
quantum fluctuations in the curvature to classical superhorizon
curvature perturbations.  These then become the
density perturbations seen in the CMB and that seed the growth
of large-scale structure in the later Universe.  We moreover see
that the primordial curvature perturbation is a realization of a
random field in which each Fourier amplitude $\calR_{\boldsymbol k}$ is
selected from a Gaussian distribution with variance,
$\VEV{|\calR_{\boldsymbol k}|^2} = H^2/(4 \Mpl^2 \epsilon k^3)$.  We
then define the curvature power spectrum,
\begin{equation}
     \Delta_{\calR}^2(k) \equiv \frac{k^3}{2\pi^2} \VEV{|\calR|^2} 
     = \frac{1}{8\pi^2} \frac{H^2}{\Mpl^2 \epsilon} \simeq
     \frac{1}{24\pi^2} \frac{V}{\Mpl^4 \epsilon},
\label{eqn:curvaturepower}
\end{equation}
the contribution per logarithmic interval in $k$ to the
real-space curvature variance $\VEV{\calR^2} = \int\, d\ln k
\Delta_{\calR}^2(k)$ (the last term is based on the relation $H^2\propto V$,
from the Friedmann equation, which is valid during slow-roll inflation). 
From current constraints,
$\Delta_{\calR}^2 \simeq 2.2\times 10^{-9}$ [now measured to
$\lesssim 1\%$ at $2\sigma$ \citep{Ade:2015xua}], we infer an
upper limit $V^{1/4} \leq 6.6\times
10^{16}\,\epsilon^{1/4}$~GeV to the energy scale of inflation.
If we assume $\epsilon \lesssim 0.1$, then this is $V^{1/4}
\lesssim 3.7\times 10^{16}$~GeV.

\subsubsection{Spectral index for primordial density
perturbations}
The spectral index $n_s(k)$ for the matter power spectrum is
determined by the logarithmic derivative of the power spectrum
with respect to wavenumber through,
\begin{equation}
     n_s(k)-1 \equiv \frac{d \ln \Delta_{\calR}^2(k)}{ d \ln k}.
\end{equation}
The scale factor $a$ varies much more rapidly than $H$ during
inflation, and we evaluate the power-spectrum amplitude at
$k=aH$.  Therefore, $d\ln k = dk/k \simeq da/a  =(\dot a/
a)dt=H\,dt$. 
From this, and using $\Delta_{\cal R}^2(k) \propto H^2/\epsilon$,
we infer $n_s-1=2\eta - 6\epsilon$ in terms of the slow-roll
parameters $\epsilon$ and $\eta$ defined above.
A spectrum with $n_s=1$, the ``Peebles-Harrison-Zelddovich''
spectrum \citep{Peebles:1970ag,Harrison:1969fb,Zeldovich:1969sb},
was postulated well before the advent of inflation
simply because this power-law index keeps the perturbation
amplitude small on large scales (to preserve the large-scale
homogeneity of the Universe) and on small scales (to preserve
the sucesses of big-bang nucleosynthesis).  Inflation then 
provided a physical mechanism for generating perturbations
with $n_s \simeq 1$.  If inflation is at work, though, then some
{\it departure} from $n_s=1$ is to be expected.  In single-field
slow-roll inflationary models, $n_s-1$ can be either positive
(in which case the spectrum is said to be ``blue'') or negative (a
``red'' spectrum).  Planck data now
indicate $n_s=0.968 \pm0.006$, a $\gtrsim5\sigma$ discrepancy
with $n_s=1$ \citep{Ade:2013zuv,Ade:2015xua}, confirming earlier
indications \citep{Komatsu:2008hk,Calabrese:2013jyk}.
The finding $n_s\neq 1$ thus supports the notion
of inflation.  If interpreted within the context of slow-roll
inflation, it places very important new constraints to the slope
and curvature of the inflaton potential by constraining
$6\epsilon -2\eta = 0.032\pm0.006$.

Note also that the power-law index is expected to run with
scale \citep{Kosowsky:1995aa}; i.e., the primordial power
spectrum is not a pure power
law.  In particular, $dn/d\ln k = -16 \epsilon \eta + 24
\epsilon^2 + \xi^2$, where $\xi^2 \equiv (m_{\rm Pl}^4/ 64
\pi^2)(V' V'''/ V^2)$.  Current constraints are
consistent with the small value for this running expected if
$\epsilon, \eta\lesssim 0.1$ \citep{Ade:2015xua}.

\subsection{Gravitational waves (tensor metric perturbations)}
\label{sec:GWs}
Maxwell's equations in the absence
of sources result in a wave equation for the electromagnetic
fields.  The propagating solutions to these equations are
electromagnetic waves, which come with two different
linear polarizations. The quanta of these waves are photons. 
Likewise, the sourceless Einstein's equations for the spacetime
metric imply propagating gravitational waves that come in two
linear polarizations, denoted $+$ and $\times$; the quanta of
these gravitational waves are gravitons.

Gravitational waves are waves in the transverse
($\partial^i h_{ij}$) and traceless ($h^i_{\,i}$) components of
the metric perturbation, defined in the FRW Universe in terms of
the spatial components of the metric, by $g_{ij} =
a^2(\delta_{ij} + 2 h_{ij})$. 
The Einstein-Hilbert action for the metric, expanded to
quadratic order $in h_{ij}$, is
\begin{equation}
     S_h = \frac14 \int\, dt\, \int\, d^3\boldsymbol x \, a^3 
     \Mpl^2 \left[ (1/2)\left(\dot h_{ij} \right)^2 - \frac{1}{2a^2}
     \left( \partial_k h_{ij} \right)^2 \right].
\end{equation}
When written in terms of Fourier modes and in terms of the two
gravitational-wave polarizations, of amplitudes $h_+$ and
$h_\times$, this becomes,
\begin{equation}
     S_\calR = \sum_{p=+,\times} \sum_{\boldsymbol k} \int\, dt\, a^3 \left[(1/2) |\dot
     v_{p,\boldsymbol k}|^2 - (1/2)(k/a)^2 |v_{p,\boldsymbol k}|^2 \right],
     \label{eq:EinHilAct}
\end{equation}
with $v_p = (\Mpl/2) h_p$. Equation \ref{eq:EinHilAct} is identical to
Equation \ref{eqn:oscillatoractions}, apart from the sum over
polarizations.  In other words, each Fourier mode and
polarization state of the gravitational wave has an amplitude
that behaves like that of a simple harmonic oscillator.
Following the same reasoning that led to
Equation \ref{eqn:curvaturepower}, we find a gravitational-wave
power spectrum (after summing over the two polarizations),
\begin{equation}
     \Delta_h^2(k) \equiv 2 \frac{k^3}{2\pi^2} \VEV{
     \left|h_{p,\boldsymbol k}\right|^2} = \frac{2}{\pi^2}
     \frac{H^2}{\Mpl^2}.
\end{equation}
Given that $H^2\propto V$ during inflation, the 
gravitational-wave amplitude is thus determined
entirely by the energy density of the Universe during inflation.

The gravitational-wave amplitude is often reported as a
tensor-to-scalar ratio,
\begin{equation}
     r \equiv \frac{\Delta_h^2}{\Delta_{\calR}^2} = 16\epsilon
     \simeq 0.1 \left( \frac{V}{\left[2\times 10^{16}\,{\rm
     GeV}\right]^4}\right),
\label{eqn:rV}
\end{equation}
where the measured value of $\Delta_{\calR}^2$ was used in the
last step.  The current bound $r\lesssim0.1$
\citep{Ade:2015tva,Ade:2015lrj} thus provides a slightly stronger
constraint to the energy density than the bound from measurement
of the scalar amplitude.

As with density perturbations, the gravitational-wave power
spectrum is $k$ independent only to the extent that the Hubble
parameter $H$ is constant during inflation.  Most 
generally, the inflaton $\phi$ rolls down the potential
$V(\phi)$, and so $H$ decreases as inflation proceeds.  There is
thus a tensor spectral index,
\begin{equation}
     n_t = \frac{d \ln \Delta_h^2(k)}{d \ln k} = -2\epsilon.
\end{equation}
which is required to be negative in single-field slow-roll
inflation; i.e., gravitational waves are said to have a ``red''
spectrum.  This arises because the energy density during
inflation is monotonically decreasing with time.

\begin{textbox}
\subsubsection{A worked example: Power-law potentials}
Here we illustrate the evaluation of the ``inflationary
observables''---the scalar and tensor power-spectrum amplitudes
and their spectral indexes---for a class of inflationary models
described by power-law potentials \citep{Linde:1983gd}.  We thus
take the inflaton potential to be
$V(\phi) = (1/2)m^{4-\alpha} \phi^\alpha$, where $m$ is a parameter
with dimensions of mass or energy, and $\alpha$ is the power-law
index.  Using the formulas derived in Section
\ref{sec:inflationbasics}, the number of $e$-foldings
of
inflation between the time that the field takes the value $\phi$
and the time its value is $\phi_{\rm end}$ at the end of inflation (when
$\epsilon(\phi)\simeq 1$) is $N\simeq (\phi^2-\phi_{\rm
end}^2)/(2\alpha \Mpl^2)$.
The slow-roll parameters are then given in terms of $N$ by
$\epsilon = \alpha/(4N)$ and $\eta=(\alpha-1)/(2N)$.  We also
have the scalar spectral index $n_s-1= - (2+\alpha)/(2N)$ and
$r=4\alpha/N$.  From the value $n_s\simeq0.968$ from Planck, we
infer, for power-law potentials, $N =
62.5\,[1+(\alpha-2)/4][(1-n_s)/0.032]^{-1}$, $\epsilon \simeq
0.008\,(\alpha/2) [1+(\alpha-2)/4]^{-1} [ (n_s-1)/0.032]$, and $r
\simeq 0.13 (\alpha/2) [1+(\alpha-2)/4]^{-1}  [ (n_s-1)/0.032]$. 
Using $\phi^2_{\rm end}\simeq \alpha^2\Mpl^2/2$ (from
$\epsilon(\phi_{\rm end}) \simeq 1$), we infer that the energy
density at the end of inflation is $V(\phi_{\rm end})=
m^{4-\alpha} (\alpha \Mpl/\sqrt{2})^\alpha/2$.  We also infer
that the inflaton must traverse a distance $\Delta \phi \gtrsim
16\, (\alpha/2)^{1/2} [1+(\alpha-2)/4]^{1/2}
[(1-n_s)/0.032]^{-1/2}\, \Mpl$ during inflation.

\end{textbox}

\subsubsection{How big is $r$?}  
Given the considerable effort required to seek inflationary
gravitational waves, it is important to ask how big the
inflationary gravitational wave (IGW)  amplitude $r$ is expected to
be.  Although there are indeed reasonable models that allow for
almost arbitrarily small $r$ (as we will illustrate below),
there are a variety of arguments that suggest a
value $r\gtrsim 10^{-3}$, accessible experimentally within the
next 5--10 years.

Within the context of SFSR inflation, the value of $r$ is
given once a potential $V(\phi)$ is specified.  As
Equation \ref{eqn:rV} indicates, the gravitational-wave amplitude
is fixed by the energy density $V$ during inflation
\citep{Starobinsky:1979ty,Rubakov:1982df,Fabbri:1983us,Abbott:1984fp}.
Originally, the inflaton was thought to have something to do
with a Higgs field associated with grand unification. If so,
then a value $V^{1/4}\sim 10^{16}$~GeV (the grand unification 
energy scale in supersymmetric theories \citep{Dimopoulos:1981yj}) was to to be expected,
leading to a gravitational-wave amplitude in the ballpark of
$r\sim0.01-0.1$.  In the years since then there have been a
plethora of ideas that identified the inflaton with new physics
associated with Peccei-Quinn symmetry breaking
\citep{Turner:1990uz}, supersymmetry
breaking \citep{Kachru:2003aw}, 
electroweak-symmetry breaking \citep{Knox:1992iy}, and
a number of other
ideas for new physics at energy scales well below that of grand
unification.  In these models, $r$ is predicted to be far
smaller---e.g., as small as $r\sim 10^{-52}$ if the energy scale
of inflation is $V^{1/4}\sim$TeV, as may occur if the inflaton
has something to do with electroweak symmetry breaking.  Thus,
until recently, the question of whether the gravitational-wave
signal was strong enough to be detected boiled down, in the
minds of many theorists, to whether inflation had to do with
grand unification.

\begin{marginnote}[3.35in]
\entry{Constraints from $n_s$}{---------------------------}
\entry{$\phi^2$ Inflation}{\\ \underline{ best-fit $r$ $|$ $3\sigma$ bound} \\ $~~~~~\;\,\,\,\,0.13\,$ $|$ $0.057$}
\vspace{0.15in}
\entry{Monodromy $\phi$}{\\ \underline{ best-fit $r$ $|$ $3\sigma$ bound} \\ $~~~~\;\,\,\,0.087\,$ $|$ $0.038$}
\vspace{0.15in}
\entry{Monodromy $\phi^{2/3}$}{\\ \underline{ best-fit $r$ $|$ $3\sigma$ bound} \\ $~~~~\;\,\,\,0.065\,$ $|$ $0.028$}
\vspace{0.15in}
\entry{$R^2$ Inflation}{\\ \underline{ best-fit $r$ $|$ $3\sigma$ bound} \\ $~~~~\;\,\,\,0.003\,$ $|$ $6\times10^{-4}$}
\vspace{0.15in}
\entry{Natural Inflation}{\begin{centering} \\ \underline{ $3\sigma$ bound} \\ $~~~~~~~~\,\;\,\,\, 0.04$ \end{centering}}
\vspace{0.15in}
\entry{Higgs-like Potential}{\begin{centering} \\ \underline{ $3\sigma$ bound} \\ $~~~~~~~~\,\;\,\,\, 0.03$ \end{centering}}
\end{marginnote}

With empirical evidence in the past few years indicating that
$n_s\neq1$, the thinking on the magnitude of $r$ has shifted.
The reasoning, which can be presented in several ways, suggests
a value $r \gtrsim 0.001$, although the arguments are never fully
conclusive.  One argument is based simply upon the relations
$n_s-1 = 2 \eta- 6\epsilon$ and $r=16\epsilon$.  If $\eta \sim
\epsilon$, then $n_s\simeq 0.968$ implies $\epsilon \sim 0.01$.
In this case a value $r\sim 0.1$ is to be expected.  If for some
reason $\eta \ll \epsilon$, then the $3\sigma$ limit $1-n_s
\gtrsim 0.014$ implies $r\gtrsim 0.037$.  

There are then lower
limits to $r$ that can be obtained within the context of any
particular class of inflaton potentials.  For example, for the
power-law potentials considered above, $1-n_s \gtrsim 0.014$
implies $r \gtrsim 0.057 (\alpha/2)[1+(\alpha-2)/4]^{-1}$.  For
$\alpha=1$ and $\alpha=2/3$ [two values that arise in
axion-monodromy models of inflation
\citep{McAllister:2008hb,Silverstein:2008sg}], the limit
evaluates to $r\gtrsim 0.038$ and $r\gtrsim 0.028$,
respectively.

Power-law potentials constitute just one of a number of
families of inflaton potentials that is far too big to review
here.  We thus instead just provide a few illustrative
examples.  Natural inflation
\citep{Adams:1992bn,Freese:2014nla} involves a potential $V(\phi)
\propto 1-\cos(\phi/v)$, where $v$ is a parameter.  The analytic
relations are not as simple as those for power-law relations;
still, one finds $r\gtrsim 0.04$ for the $3\sigma$ upper limit
$1-n_s\lesssim 0.05$ \citep{Munoz:2014eqa}.  Inflation with a 
Higgs-like potential, $V(\phi) \propto (\phi^2-v^2)^2$, where $v$ is a
parameter \citep{Kaplan:2003aj}, requires $r\gtrsim0.03$ for $1-n_s \lesssim 0.05$
\citep{Munoz:2014eqa}.
One can consider a scenario where $\epsilon \ll \eta$, in
which case $r$ could be far smaller than in the hitherto
considered example.  The unusually small slope $V'$ this
scenario requires can be arranged if the inflaton is near a
local maximum of $V(\phi)$.  Given that $V'$ changes as the
inflaton rolls away from the maximum, this scenario requires
some tuning, and simple
implementations of this idea (for example, in Higgs-like
inflation, where the field begins near a local maximum of the
potential) still result in values $r\gtrsim 0.01$.
Alternatively, a low value of $r$ can be generated by
inflection-point inflation \citep{Itzhaki:2007nk},  wherein the
inflaton happens to occupy a point in a potential where both
$V'$ and $V''$ are very small.

\begin{figure}
\includegraphics[width=\linewidth]{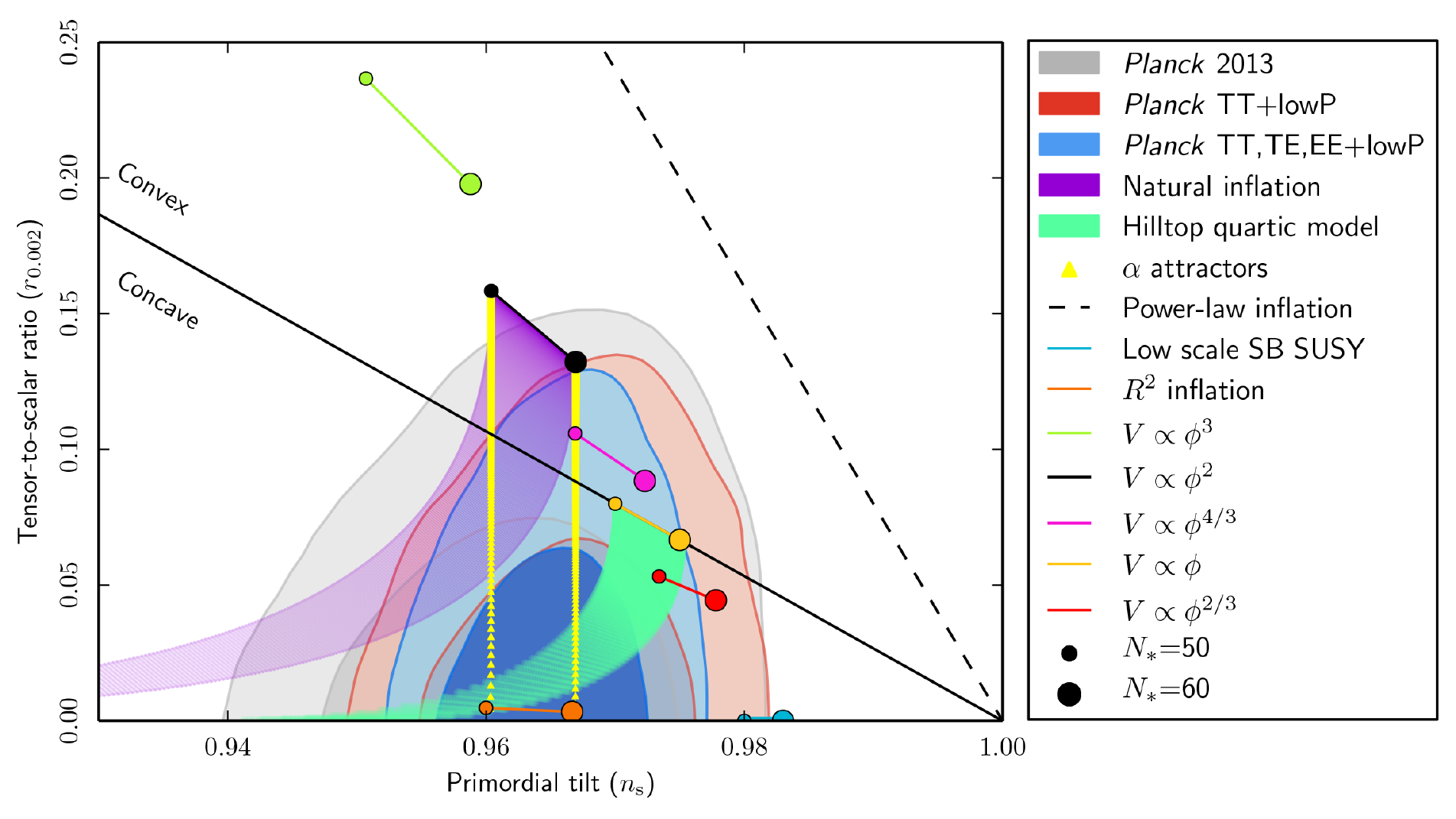}
\caption{ The current constraints, from the Planck Collaboration
     \citep{Ade:2015lrj}, to the scalar spectral index $n_s$
     and tensor-to-scalar ratio $r$.  Shown are predictions for
     a variety of slow-roll models.  The ``concave'' and
     ``convex'' refer to the sign of the second derivative
     $V''$ of the inflaton potential.}
\label{fig:nsr}
\end{figure}

Other small-$r$ potentials arise from inflation models based on
alternative gravity.  For example, Starobinsky's $R^2$ model
\citep{Starobinsky:1980te,Mukhanov:1981xt,Ellis:2013xoa,Buchmuller:2013zfa}, in which the
action for gravity contains a term quadratic in the Ricci scalar
$R$ in addition to the linear term that appears in the usual
Einstein-Hilbert action.  The resulting dynamics
can be mapped onto those of SFSR
inflation with a fairly exotic-looking potential.  This model
predicts $r = 3 (n_s-1)^2$ which yields $r\simeq 0.003$ for the
Planck central value of $n_s\simeq 0.968$ and $r\gtrsim 6\times
10^{-4}$ for the $3\sigma$ limit $1-n_s \gtrsim 0.014$. The same 
prediction holds for the ``Higgs Inflation" model \citep{Bezrukov:2007ep}.
Several potentials [e.g., a Coleman-Weinberg potential
\citep{Coleman:1973jx} and others \citep{Kinney:1995cc}]
considered prior to the discovery $n_s\simeq1$ to illustrate
that $r$ could be virtually arbitrarily small predict, with
current constraints to $n_s$, values $r\gtrsim 0.01$
\citep{Kinney:private}.  Although
this does not rule out the possibility  $r\ll 0.001$ in SFSR, it
indicates the increased pressure on low-$r$ SFSR models provided
by the measurement of $n_s$.

The estimates done here (and listed in the sidebar), which have assumed a specific $3\sigma$ error range for $n_s$, actually simplify a bit the actual current constraints to the model parameter space.  As shown in {\bf Figure \ref{fig:nsr}}, experiments provide joint constraints to the $n_s$-$r$ parameter space \citep{Ade:2015lrj}, and so the constraints to models are generally stronger than what we have assumed here.

\paragraph{Large-field versus small-field models}
There is an
interesting model-independent argument that suggests a qualitative
difference between SFSR models with $r\gtrsim 10^{-3}$ and those
with $r\lesssim 10^{-3}$ \citep{Turner:1993su,Lyth:1996im}.  This 
``Lyth bound'' follows from Equation \ref{eq:amount} and the relation 
$r\simeq16\,\epsilon$. If $\epsilon$ is roughly constant during
inflation, then we infer that the distance $\Delta \phi$
traversed by the inflaton during inflation is
\begin{equation}
     \frac{\Delta \phi}{\Mpl} \gtrsim \sqrt{\frac{r}{8}}N \simeq
     \left(\frac{r}{10^{-3}} \right)^{1/2}\left(\frac{N}{10}\right),
\end{equation}
where the inequality comes from the fact that there may be many
more $e$-folds of inflation that precede those required for
observations within our horizon.  Thus, $r\gtrsim10^{-3}$
requires $\Delta \phi>\Mpl$, a ``large-field'' model of
inflation, while $r\lesssim 10^{-3}$ allows for $\Delta \phi
\lesssim \Mpl$ (a ``small-field'' model).

While the construction of any workable SFSR potential requires
what virtually any particle theorist would consider fine
tuning (in order to sustain slow-roll for the required number of 
e-folds), large-field models pose an even greater challenge for model builders.  
Suppose we Taylor expand the potential about its minimum (assumed to be
$V=0$),
\begin{equation}
     V(\phi) = \frac{1}{2} m^2\phi^2 + \phi^2 \sum_{p=1}^\infty
     \lambda_p \left( \frac{\phi}{\Mpl}\right)^p.
\end{equation}
We can always choose the coefficients $\lambda_p$ so that
$\epsilon,\eta \ll1$ at some particular value of $\phi$.  These
coefficients, though, are expected to receive contributions
$\Delta \lambda_p(\phi)$ from quantum corrections that are
themselves functions of $\phi$. These corrections are then expected to vary by
order unity over distances $\Delta \phi \gtrsim \Mpl$.  It is
thus difficult to see how a generic potential can preserve
$\epsilon,\eta \ll1 $ over an inflaton displacement $\Delta \phi
\gtrsim \Mpl$. This problem is averted if there are
symmetries that set some of these coefficients or the corrections
to zero, and a theoretical industry has developed to construct
string-inspired models that preserve $\epsilon,\eta \ll1$ with
large field excursions \citep{Baumann:2009ni,Baumann:2014nda}.
It should also be emphasized that the boundary between
small-field and large-field models is blurry \citep{Itzhaki:2008hs}. 
Still, an experiment with a detection sensitivity of $r\sim 10^{-3}$ 
would provide a fairly definitive statement about the validity 
(or otherwise) of large-field SFSR models.

\paragraph{Beyond-SFSR inflation}
As will be discussed below, the
predictions of SFSR inflation are in exquisite agreement with a
wealth of precise measurements, and there are no experimental
nor observational results that drive us to introduce any new
physics beyond that found in SFSR inflation.  Still, SFSR
inflation should be viewed as no more than a working or toy
model, and a vast theoretical literature that explores ideas for
beyond-SFSR inflation has evolved.  In many of these, the
connections between $n_s$ and $r$ found in SFSR models are
either revised or lost.  For example, in models with
modifications to the kinetic term in the inflaton Lagrangian
\citep{Alishahiha:2004eh,ArmendarizPicon:1999rj},
the GW amplitude can be suppressed by the speed of sound $c_s$,
for inflaton perturbations, which may be very small.  Even so,
the remarkable success of the simplest SFSR model, along with
current constraints to $n_s$, make a compelling case, many theorists
would agree, for a measurement sensitive to $r\sim0.001$.

In this context, it is worth mentioning that an important property of 
simple models of inflation---that a measurement of the amplitude 
of tensor modes is in one-to-one correspondence with the energy
scale of inflation---has been augmented in models 
involving spectator fields that are excited from the vacuum during 
inflation and possibly enhance the tensor fluctuation spectrum 
\citep{Senatore:2011sp, Cook:2011hg, Barnaby:2012xt}. However, 
these models appear to require complex setups \citep{Carney:2012pk, Kleban:2015daa}, 
and tend to produce tension with constraints on scalar fluctuations and 
non-gaussianity \citep{Mirbabayi:2014jqa, Ozsoy:2014sba}.

\section{From gravitational waves to the CMB}
\label{sec:GWstoQU}

We now show how gravitational waves induce
temperature fluctuations and polarization in the cosmic
microwave background.  Following the pioneering work
of \citet{Polnarev} [see also
\citet{Cabella:2004mk}], we first derive the angular
distribution of photon intensities in the presence of a
gravitational wave (GW).  Suppose that the Universe is
filled with photons that do not scatter. In this case, the
photon energies are affected only by the form of the metric.
Consider a single monochromatic plane-wave gravitational wave,
which appears as a tensor perturbation to the FRW metric,
\begin{equation}
     ds^2=a^2(\eta)\left[d\eta^2-dx^2(1+h_+)+dy^2(1-h_+)+dz^2\right],
\label{eqn:perturbedmetric}
\end{equation}
where $\eta$ is the conformal time and
\begin{equation}
     h_+(\boldsymbol{x},\eta) \simeq h(\eta)e^{ik\eta}e^{-ikz},
\label{eqn:hevolution}
\end{equation}
describes a plane wave propagating in the $\hat{z}$
direction. This is a linearly-polarized gravitational wave with
``+'' (rather than ``$\times$'') polarization.  Here $h(\eta)$ is the
amplitude; at early times when $k\eta\lesssim 1$,
$h(\eta)\simeq{\rm const}$, but then $h(\eta)$ redshifts away when
$k\eta\gtrsim 1$.  If we construct the Einstein tensor
$G_{\mu\nu}$ from the metric, Equation \ref{eqn:perturbedmetric},
then the vacuum Einstein equation $G_{\mu\nu}=0$ leads to the wave equation for
$h_+(\boldsymbol x,t)$.  The ``$\simeq$'' sign appears in
Equation \ref{eqn:hevolution} because the gravitational waves do
not propagate in a vacuum but rather in a Universe filled with a
cosmic fluid.  The anisotropic stress of this fluid (to which
the neutrino background contributes after neutrinos decouple)
modifies slightly the time evolution, a calculable $\sim10\%$
correction to Equation \ref{eqn:hevolution}
\citep{Weinberg:2003ur,Pritchard:2004qp,Dicus:2005rh}.

Photons that propagate freely through this spacetime experience
a frequency shift $d\nu$ during an expansion interval $d\eta$ determined
by the geodesic equation, which in this spacetime takes the form,
\begin{equation}
     \frac{1}{\nu}\frac{d\nu}{d\eta}=-\frac{1}{2}(1-\mu^2)\cos2\phi
     e^{-ikz}\frac{d}{d\eta}(he^{ik\eta}),
\end{equation}
where $\mu$ is the cosine of the angle that the photon trajectory
makes with the $z$ axis, and $\phi$ is the azimuthal angle of
the photon's trajectory.  This redshifting is
polarization-independent, but polarization is then induced by
Thomson scattering of this anisotropic radiation field.  To
account for the polarization, we must follow the time evolution
of four distribution functions (DFs) $f_s(\boldsymbol x,\boldsymbol q;\eta)$
\citep{Crittendenthesis,Crittenden:1993wm,Crittenden:1994ej}, where
$\boldsymbol q$ is the photon momentum, for $s=I$, $Q$, $U$, and $V$,
the four Stokes parameters required to specify the
polarization.  The original (unperturbed) distribution function
is $\bar f_I(\boldsymbol q,\boldsymbol x; \eta) = \left[e^{h
\nu/k_B T(\eta)}-1 \right]^{-1}$, where $k_B$ is the Boltzmann constant
and $T(\eta)$ the unperturbed CMB temperature at conformal time
$\eta$, and $\bar f_Q=\bar f_U=\bar f_V=0$.   We then define
perturbations $\Delta_s e^{i \boldsymbol k \cdot \boldsymbol x}= 4 \delta
f_s/(\partial \bar f/\partial \ln T)$, suppressing an index $\boldsymbol{k}$ 
for notational economy.  Thomson scattering
induces no ciruclar polarization,
so $\Delta_V=0$ at all times.  Since the gravitational redshift
and Thomson scattering are frequency independent, the
evolution of the distribution function is the same for all
frequencies.  Since the $e^{i \boldsymbol k \cdot \boldsymbol x}$ spatial
dependence of the DFs is separated out in the definition of
$\Delta_s$, the perturbed DFs are functions $\Delta_s(\hat q;
\eta)$ only of the direction $\hat q$ of the photon and the
conformal time $\eta$.  Finally, if we
define perturbation variables $\tilde \Delta_s$ by
\begin{equation}
     \Delta_I= \tilde \Delta_I(1-\mu)^2 \cos 2\phi,
    ~~~~~~
     \Delta_Q = \tilde \Delta_Q(1+\mu)^2 \cos 2\phi,
  ~~~~~~
     \Delta_U = \tilde \Delta_U 2\mu \sin 2\phi,
\label{eqn:redefinitions}
\end{equation}
the new variables $\tilde\Delta_s(\mu;\eta)$ are
now functions only of $\mu$ and there is a relation $\tilde
\Delta_Q = -\tilde \Delta_U$ for the gravitational wave, a
consequence of the fact that the orientation of the photon polarization
is fixed by the direction of the photon with respect to the GW
polarization tensor.  As a result, the Boltzmann equations for
the distribution functions reduce to two equations
\citep{Crittenden:1993wm,Kosowsky:1994cy,Ma:1995ey},
\begin{equation}
     \tilde{ \dot \Delta}_I + i k \mu \tilde \Delta_T = -\dot h -
     \dot \kappa \left[ \tilde \Delta T - \Psi \right],
 ~~~~~~~     \tilde {\dot \Delta}_Q + i k \mu \tilde \Delta_Q = -\dot
     \kappa \left[\Delta_P + \Psi \right],
\label{eqn:Boltzmanns}
\end{equation}
where the dot denotes derivative with respect to conformal
time.  Here, the variable
\begin{equation}
     \Psi \equiv \left[ \frac{1}{10}\tilde \Delta_{I0} + \frac17
     \tilde \Delta_{T2} + \frac{3}{70} \tilde \Delta_{T4}
     -\frac35 \tilde \Delta_{Q0} + \frac67 \tilde \Delta_{Q2} -
     \frac{3}{70} \tilde \Delta_{Q4} \right],
\label{eqn:Psi}
\end{equation}
is given in terms of the Legendre moments $\tilde \Delta_{I\ell}(\eta)
=(1/2) \int_{-1}^1 \, d\mu\, \tilde \Delta_I(\mu;\eta)
P_\ell(\mu)$ (and similarly for $\tilde \Delta_{Q\ell}$), where
$P_\ell(\mu)$ is a Legendre polynomial.  The quantity
$\dot\kappa\delta = (d\kappa/d\eta)d\eta $ is the contribution to
the Thomson optical depth in the conformal-time interval $d\eta$.

Equations \ref{eqn:Boltzmanns} and \ref{eqn:Psi} look complicated
but describe relatively simple physics. The left-hand sides of
Equation \ref{eqn:Boltzmanns} are simply the Lagrangian time
derivatives for a Fourier mode of wavenumber $k$.  The $\dot h$
in the first equation accounts for the intensity variation
(described above) induced by the gravitational redshift; its
absence from the second equation is because the gravitational
redshift is polarization-independent.  As the presence of the
differential Thomson optical depth $\dot\kappa$ suggests, the
terms on the right-hand sides of Equations \ref{eqn:Boltzmanns}
involving $\Psi$, $\tilde\Delta_I$, and $\tilde \Delta_P$
account for Thomson scattering.  They are derived using the
dependence $d\sigma_T/d\Omega \propto (\hat \epsilon_i \cdot
\hat \epsilon_f)^2$ of the Thomson differential cross section on
the polarization vectors $\hat \epsilon_i$ and $\hat \epsilon_f$
of the initial- and final-state photons. This dependence also 
explains why a quadrupolar anisotropy in the incoming radiation 
is required in order to generate the linear polarization signal,
as was first realized in \citet{1968ApJ...153L...1R}.

Still, Equations \ref{eqn:Boltzmanns}) and \ref{eqn:Psi}
constitute a set of coupled partial integro-differential
equations.  In practice, they are solved numerically by
expanding $\tilde\Delta_I$ and $\tilde \Delta_Q$ in terms of
their Legendre moments and thus recasting the equations as an
infinite set of coupled Boltzmann equations for
$\tilde\Delta_{I\ell}(\eta)$ and $\tilde \Delta_{Q \ell}(\eta)$
\citep{Crittenden:1993wm,Kosowsky:1994cy}.
They are then solved numerically by integrating from some early
time and truncating the hierarchy at some sufficiently high
$\ell$.  

\begin{figure}
\includegraphics[width=\linewidth]{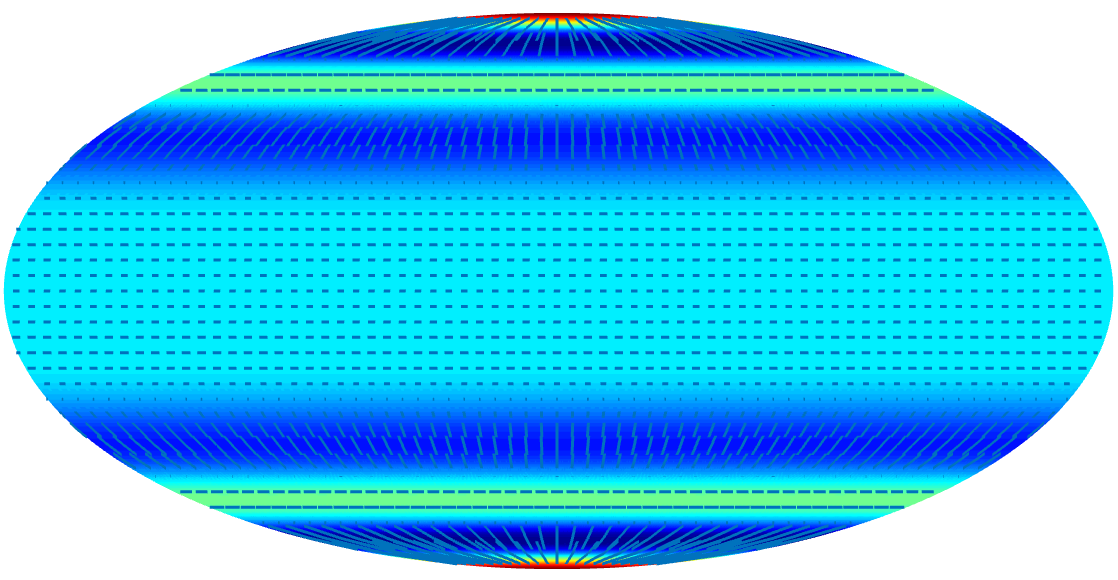}
\includegraphics[width=\linewidth]{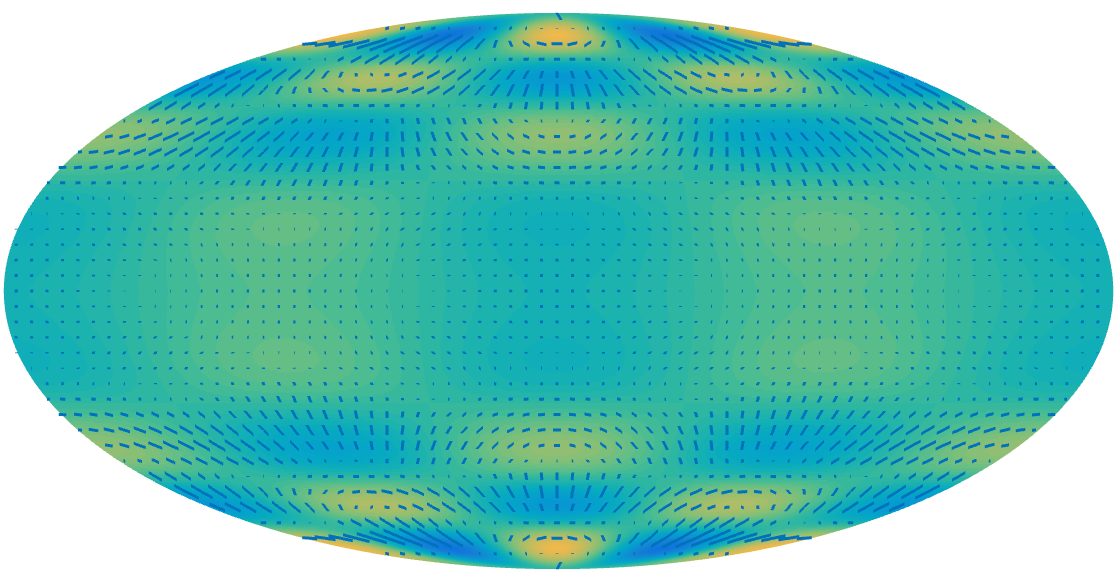}
\caption{{\it Top:} The CMB temperature-polarization pattern induced by one
     Fourier mode of the density field (i.e., a scalar metric
     perturbation).  The polarization pattern varies along a
     direction parallel/perpendicular to lines of constant
     longitude that align with the direction of the wave.  The
     induced polarization pattern is thus a pure E mode. {\it Bottom:}
     The same for a single gravitational wave (i.e., a single
     Fourier mode of the tensor field).  We see that in this
     case, there is variation of the polarization not only
     parallel/perpendicular to lines of constant longitude, but
     also along directions $45^\circ$ with respect to these
     lines.  There is thus a B mode induced.}
\label{fig:onewave}
\end{figure}

Later we will return to these equations, but for now we 
show in {\bf Figure \ref{fig:onewave}(b)} the resulting CMB
temperature-polarization pattern induced by one gravitational
wave propagating in the $\hat z$ direction 
\citep{Caldwell:1998aa}.  The quadrupolar
variation (i.e., the $\cos2\phi$ dependence) of the
temperature-polarization pattern can be seen as one travels
along a curve of constant latitude, and the wavelike pattern
can be seen as one moves along a line of constant longitude.  It
can be seen that as we move along the line of constant longitude, 
there are variations in $Q$, the component of the
polarization perpendicular/parallel to those constant-longitude
lines.  It can also be seen, however, that there are variations
in $U$, the component of the polarization $45^\circ$ with
respect to constant-longitude lines.  This, as we will see
below, is a signature of the B mode in the CMB polarization pattern induced by
the gravitational wave.  This is to be contrasted with the
polarization pattern, shown in {\bf Figure \ref{fig:onewave}(a)}, for a
single Fourier mode of the density field.  In this case, there
is no variation along lines of constant latitude, and there is
only variation in $Q$, and thus no B mode, as we will see.

Still, inflation predicts not a single gravitational wave of
given wavevector and polarization, but rather a statistically
isotropic stochastic background of gravitational waves.  The
next step is thus to understand how to represent the
polarization pattern induced by this stochastic background.

\section{Harmonic analysis for CMB polarization}

We therefore turn to the mathematical description of the polarization.
Stokes parameters $Q$ and $U$ are coordinate-dependent
quantities.  Suppose that they are measured with respect to some
$x$-$y$ axes and that we then consider some other $x'$-$y'$
axes rotated by an angle $\alpha$ with respect to the $x$-$y$
axes.  Under this rotation, the Stokes parameters $(Q,U)$
transform as components of a  symmetric trace-free (STF)
$2\times 2$ tensor,
\begin{equation}
     \left( \begin{array}{cc} Q& U\\ U & -Q\\ \end{array} \right
     ) \Rightarrow \left( \begin{array}{cc} \cos\alpha &
     \sin\alpha\\ -\sin\alpha & \cos\alpha \\ \end{array}
     \right) \left( \begin{array}{cc} Q& U\\ U &
     -Q\\ \end{array} \right ) \left( \begin{array}{cc}
     \cos\alpha & -\sin\alpha\\ \sin\alpha & \cos\alpha
     \\ \end{array} \right).
     \label{eq:UQTrans}
\end{equation}
Alternatively and equivalently, if we represent the
polarization by a complex number $P=Q+iU$, then $P\rightarrow
Pe^{2i\alpha}$ under a rotation of the coordinate axes by an
angle $\alpha$; i.e., the polarization is a spin-2 field.

Anything we say about Stokes parameters $Q$ and $U$ are
thus tied to the coordinate system we choose.  We will therefore
want to find a coordinate-system--independent representation of
this tensor field if we are to make statements about physics
that are independent of coordinate system.  Later, we will
do this on the full sky, but we first
do the simpler case of a flat sky (which also serves as a good
approximation to a small region of the sky).

\subsection{Harmonic analysis on a flat sky}
\label{sec:smallsky}

Once the polarization, $Q(\boldsymbol{\theta})$ and
$U(\boldsymbol{\theta})$, has been measured as a function of position
$\boldsymbol{\theta}=(\theta_x,\theta_y)$ on a flat region of sky, we
have measured the polarization tensor field,
\begin{equation}
     {\cal P}_{ab} = \frac{1}{\sqrt{2}}\left( \begin{array}{cc}
     Q(\boldsymbol{\theta})& U(\boldsymbol{\theta})\\ U(\boldsymbol{\theta}) &
     -Q(\boldsymbol{\theta})\\ \end{array}
\right ),
\label{eqn:QUmatrix}
\end{equation}
where the normalization is chosen so that ${\cal }{\cal P}^{ab} {\cal
P}_{ab}=Q^2+U^2$
and so that the conventions for the E and B modes defined below
agree with those of \citet{Seljak:1996gy} and \citet{Zaldarriaga:1996xe}
and are identified with the G and C modes of
\citet{Kamionkowski:1996zd,Kamionkowski:1996ks} through
$a_{(\ell m)}^{\rm G}= a_{\ell m}^{\rm E}/\sqrt{2}$ and $a_{(\ell m)}^{\rm
C}= a_{\ell m}^{\rm B}/\sqrt{2}$.

We now define gradient (``E modes'') and curl (``B modes'')
components of the tensor field that are independent of the
orientation of the $x$-$y$ axes by
\begin{equation}
\label{eq:nabla2}
     \nabla^2 E=\partial_a \partial_b {\cal P}_{ab}, \qquad
     \nabla^2 B
     = \epsilon_{ac}\partial_b\partial_c {\cal P}_{ab},
\end{equation}
where $\epsilon_{ab}$ is the antisymmetric tensor. 

\begin{figure}
\includegraphics[width=\linewidth]{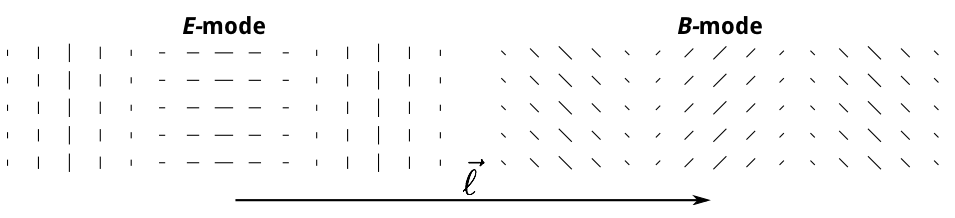}
\caption{Shown are the polarization pattern associated with a
     single E mode (left) and a single B mode (right) with
     a horizontal wavevector $\boldsymbol \ell$.  The E mode features a
     variation of the polarization along directions
     parallel/perpendicular to the direction of $\boldsymbol \ell$ (Stokes
     parameter $Q$ in a coordinate system aligned with $\boldsymbol
     \ell$), while in the B mode the variation in the polarization
     is along directions $45^\circ$ with respect to $\boldsymbol \ell$
     (Stokes parameter $U$ in coordinates aligned with
     $\boldsymbol \ell$). (From C. Bischoff.)}
\label{fig:singleEB}
\end{figure}

Writing,
\begin{equation}
     {\cal P}_{ab}(\boldsymbol{\theta}) = \int\frac{d^2\boldsymbol{\ell}}{(2\pi)^2}
     \tilde{\cal P}_{ab}(\boldsymbol{\ell}) e^{-i\boldsymbol{\ell}\cdot\boldsymbol{\theta}},
~~~~~~
\tilde{\cal P}_{ab}(\boldsymbol{\ell})=\int
     d^2\boldsymbol{\theta}{\cal P}_{ab}(\boldsymbol{\theta})e^{i\boldsymbol{\ell}\cdot\boldsymbol{\theta}},
\end{equation}
the Fourier components of $E(\boldsymbol{\theta})$ and $B
(\boldsymbol{\theta})$ are
\begin{equation}
\renewcommand\arraystretch{1.2}
\left( \begin{array}{c}  \tilde E(\boldsymbol{\ell})  \\ \tilde B(\boldsymbol{\ell}) \end{array}\right)
=   \frac{1}{\sqrt{2}} \left( \begin{array}{cc}
     \cos2\varphi_{\boldsymbol\ell} & \sin2\varphi_{\boldsymbol\ell} \\ 
     - \sin2\varphi_{\boldsymbol\ell} & \cos2\varphi_{\boldsymbol \ell}
     \\ \end{array} \right)
 \left( \begin{array}{c}  \tilde Q(\boldsymbol{\ell})  \\ \tilde U(\boldsymbol{\ell}) \end{array}\right),
\label{eqn:GCFouriercomponents}
\end{equation}
where $\varphi_{\boldsymbol\ell}$ is the angle $\boldsymbol \ell$ makes with
the $\hat x$ axis.  This relation can be inverted,
\begin{equation}
 \left( \begin{array}{c}   \tilde Q(\boldsymbol{\ell})  \\  \tilde
 U(\boldsymbol{\ell}) \end{array}\right)
=    \sqrt{2} \left( \begin{array}{cc}
     \cos2\varphi_{\boldsymbol\ell}  & -\sin 2\varphi_{\boldsymbol\ell}
     \\ \sin 2\varphi_{\boldsymbol\ell}  & \cos 2\varphi_{\boldsymbol\ell} 
     \\ \end{array} \right) 
     \left( \begin{array}{c}   E(\boldsymbol{\ell})
     \\  B(\boldsymbol{\ell}) \end{array}\right).
\label{eqn:EBtoQUFouriercomponents}
\end{equation}
Thus, for a pure B mode in the $\hat x$ direction
($\varphi_{\boldsymbol \ell}=0$), we have (as shown in the right panel
of {\bf Figure \ref{fig:singleEB}}) $\tilde Q(\boldsymbol \ell)=0$ and $\tilde
U(\boldsymbol \ell)=\tilde B(\boldsymbol \ell)$.  For a pure E mode in the
$\hat x$ direction, we have (as shown in the left panel of
{\bf Figure \ref{fig:singleEB}}) $\tilde Q(\boldsymbol \ell)=\tilde E(\boldsymbol\ell)$
and $\tilde U(\boldsymbol \ell)=0$.  Thus, {\it in an E mode, the
polarization varies parallel/perpendicular to the direction
of the Fourier mode, while for a B mode the polarization
varies along directions $45^\circ$ with respect to the direction
of the Fourier mode}.

Since the combined temperature/polarization map is described by
three sets, $\tilde T(\boldsymbol \ell)$, $\tilde E(\boldsymbol\ell)$, and
$\tilde B(\boldsymbol\ell)$, of Fourier components, the two-point
statistics of the temperature/polarization field are determined
by a total of six power spectra, $C_\ell^{{\rm X}_1,{\rm X}_2}$,
defined by
\begin{equation}
     \VEV{\tilde{X}_1(\boldsymbol{\ell})\tilde{X}_2(\boldsymbol{\ell'})} = (2\pi)^2
     \delta(\boldsymbol{\ell}+\boldsymbol{\ell'})C_\ell^{X_1X_2},
\end{equation}
where ${\rm X}_1, {\rm X}_2 =\{ {\rm T},{\rm  E}, {\rm B}\}$.
Here the angle brackets denote an average over all
realizations of the temperature map. 

Now suppose we have a given temperature/polarization map and
then consider a parity inversion; e.g., a reflection
about the $x$-axis. Then 
\begin{equation}
     \theta_y\rightarrow-\theta_y,\quad Q\rightarrow Q,\quad U\rightarrow
     -U,\quad \ell_x\rightarrow \ell_x, \quad \ell_y\rightarrow -\ell_y.
\end{equation}
Also,
\begin{equation}
     \tilde{T}(\boldsymbol{\ell})\rightarrow\tilde{T}(\boldsymbol{\ell}),\:\:\:
     \tilde E(\boldsymbol{\ell})\rightarrow\tilde E(\boldsymbol{\ell}),\:\:\:
     \tilde B(\boldsymbol{\ell})\rightarrow -\tilde B(\boldsymbol{\ell}).
\end{equation}
Thus, T and E have the same parity, while B has the
opposite parity. If the physics that gives rise to
temperature/polarization fluctuations is parity conserving, we
then expect $C_\ell^{\rm TB}=C_\ell^{\rm EB}=0$.
In this case, the statistics of the temperature/polarization map
are determined entirely by the four power spectra, $ C_\ell^{\rm
TT}$, $C_\ell^{\rm TE}$, $C_\ell^{\rm EE}$, and $C_\ell^{\rm BB}$.

\subsection{Harmonic analysis on the full sky}

If our maps extend beyond a small region of the sky, we will
have to deal with the curvature of the sky.
We thus generalize the tensor Fourier analysis 
that we carried out above for STF $2\times 2$ tensors to tensors that
live on the 2-sphere. Our discussion follows
\citet{Kamionkowski:1996ks}; a different but equivalent formalism is
presented in \citet{Zaldarriaga:1996xe}.  In the usual spherical polar
coordinates $\theta,\phi$, the sphere has a metric, $g_{ab}=
{\rm diag}(1,\sin^2\theta)$.
The polarization tensor ${\cal P}_{ab}$ must be symmetric ${\cal
P}_{ab}={\cal P}_{ba}$ and
trace-free $g^{ab}{\cal P}_{ab}=0 $, from which it follows that,
\begin{equation}
     {\cal P}_{ab}(\hat{n})= \frac{1}{\sqrt{2}}
     \left( \begin{array}{cc} Q(\hat{n}) &
     U(\hat{n})\sin\theta\\ U(\hat{n})\sin\theta &
     -Q(\hat{n})\sin^2\theta\\ \end{array} \right),
\end{equation}
where the factors of
$\sin\theta$ follow from the fact that the coordinate basis
$(\theta,\phi)$ is orthogonal but not orthonormal.

We use a colon (:) to denote a covariant
derivative on the surface of the sphere (e.g., $S^a{}_{:a}$
denotes the divergence of $S^a$) and a comma to denote a
partial derivative [e.g., $S_{,a}=(\partial S/\partial
x^\alpha)$].  Appendix A of \citet{Kamionkowski:1996ks}
reviews the rules of differential
geometry on the sphere in the notation we use here.  

Any STF $2\times2$ tensor field on the sphere can be written as
the `gradient',  $E_{:ab}-\frac{1} {2} g_{ab}E^{:c}_c$,
of some scalar field $E(\theta,\phi)$, plus the `curl,'
$(1/2) \left(B_{:ac}\epsilon^c_b+B_{:bc}\epsilon^c_{a} \right)$,
of some other scalar field $B(\theta,\phi)$,
For comparison, a vector field is analogously decomposed as
$V_a=\nabla_a E+\epsilon_{ab}\nabla_b B$.
Since any scalar field on the sphere can be expanded in
spherical harmonics (e.g. for the temperature),
\begin{equation}
     {T(\hat{n}) \over T_0}=1+\sum_{\ell=1}^\infty\sum_{m=-\ell}^\ell
     a^{\rm T}_{\ell m}\,Y_{\ell m}(\hat{n}),
     \qquad {\rm where} \qquad      a^{\rm T}_{\ell m}={1\over
     T_0}\int d\hat{n}\,T(\hat{n})   Y_{\ell m}^*(\hat{n}), 
\label{Texpansion}
\end{equation}
it follows that the polarization tensor can be expanded
in terms of basis functions that are gradients and curls of
spherical harmonics,
\begin{equation}
     {\cal P}_{ab}(\hat{n})= T_0 \sum_{\ell=2}^\infty\sum_{m=-\ell}^\ell
     \left[ a_{\ell m}^{\rm E}
     Y_{(\ell m)ab}^{\rm E}(\hat{n}) + a_{\ell m}^{\rm B} Y_{(\ell m)ab}^{\rm B}
     (\hat{n}) \right].
\label{Pexpansion}
\end{equation}
The expansion coefficients are given by
\begin{equation}
     a_{\ell m}^{\rm E}={1\over T_0}\int \, d\hat{n} {\cal P}_{ab}(\hat{n})
                             Y_{(\ell m)}^{{\rm E} \,ab\, *}(\hat{n}), \qquad\qquad
     a_{\ell m}^{\rm B}={1\over T_0}\int d\hat{n}\, {\cal P}_{ab}(\hat{n})
                                      Y_{(\ell m)}^{{\rm B} \, ab\, *}(\hat{n}),
\label{defmoments}
\end{equation}
and
\begin{equation}
     Y_{(\ell m)ab}^{\rm E} = N_\ell
     \left( Y_{(\ell m):ab} - {1\over2} g_{ab} Y_{(\ell m):c}{}^c \right),
~~     Y_{(\ell m)ab}^{\rm B} = { N_\ell \over 2}
     \left(\vphantom{1\over 2}
       Y_{(\ell m):ac} \epsilon^c{}_b +Y_{(\ell m):bc} \epsilon^c{}_a \right),
\label{Ydefn}
\end{equation}
constitute a complete orthonormal set of basis functions for the
E and B components of the polarization.  The quantity, 
$N_\ell \equiv \sqrt{ {2 (l-2)! / (l+2)!}}$,
is a normalization factor chosen so that
\begin{equation}
      \int d\hat{n}\,Y_{(\ell m)ab}^{{\rm X}\,*}(\hat{n})\,Y_{(l'm')}^{{\rm
      X'}\,\,ab}(\hat{n})
      =\delta_{\ell \ell'} \delta_{mm'},
\label{norms}
\end{equation}
for ${\rm XX}' =$EE, EB, and BB.  
Also, we can integrate by
parts to write alternatively,
\begin{equation}
     a_{\ell m}^{\rm E} = {N_\ell\over T_0}
     \int d\hat{n} \, Y_{\ell m}^*(\hat{n})\,
     {\cal P}_{ab}{}^{:ab}(\hat{n}),
\qquad\qquad
     a_{\ell m}^{\rm B} = {N_\ell\over T_0}
     \int d\hat{n} \, Y_{(\ell m}^*(\hat{n})\,
     {\cal P}_{ab}{}^{:ac}(\hat{n}) \epsilon_c{}^b.
\label{GCmomentseasy}
\end{equation}
Finally, since $T$, $Q$, and $U$ are real, we get $a_{\ell m}^{\rm X\,*} =
     (-1)^m a_{\ell,-m}^{\rm X}$, where ${\rm X}= \{{\rm T,E,B}\}$.
The temperature/polarization power spectra are now
\begin{equation}
     \VEV{a_{\ell m}^{\rm X\,*}a_{\ell'm'}^{\rm X'}} = C_\ell^{\rm X\rm
     X'}\delta_{\ell \ell'}\delta_{mm'},
\end{equation}
for ${\rm XX}' =$TT, EE, BB, TE, TB, and EB.
The $C_\ell$ here reduce in
the small-angle (large-$\ell$) limit with those in
Section \ref{sec:smallsky} as long as the angles in the
flat-sky limit are given in radians.

\begin{figure}
\includegraphics[width=\linewidth]{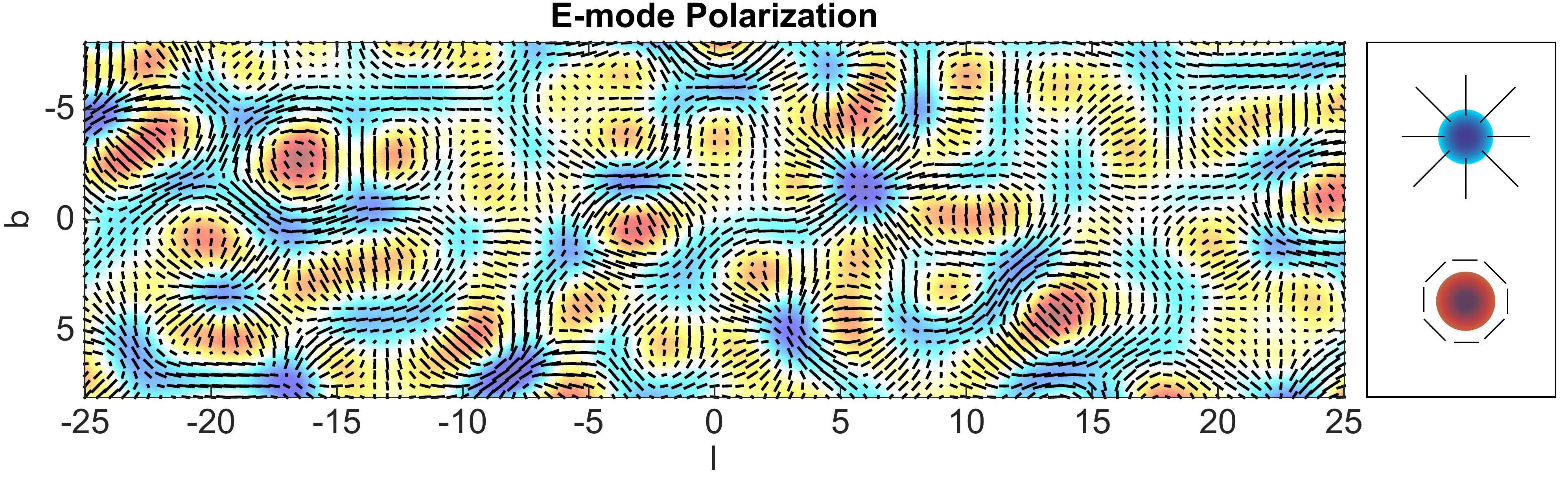}
\includegraphics[width=\linewidth]{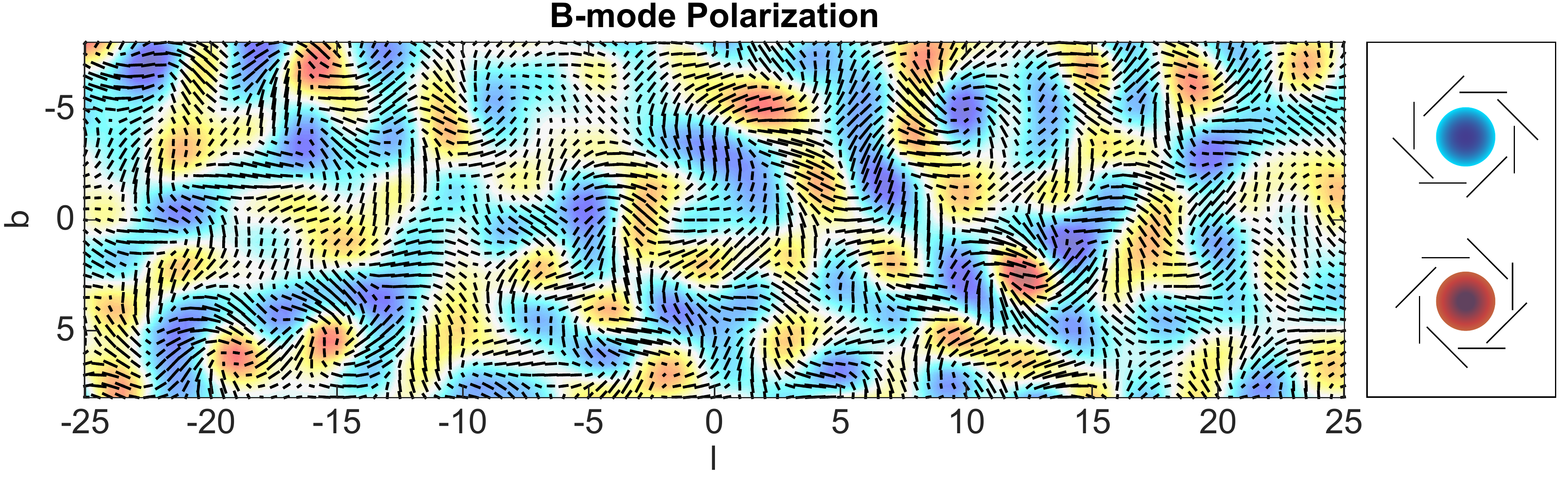}
\caption{In the top figure we show a polarization pattern composed only of E modes and in the bottom one composed only of B modes.  As indicated on the right, it is seen that around hot spots (red) the polarization pattern of the E mode is tangential and radial around cold spots (blue).  The polarization pattern surrounding hot and cold spots of the B mode show a characteristic swirling pattern (with different orientation around hot and cold spots).}
\label{fig:pureGC}
\end{figure}

The $ Y_{(\ell m)ab}^{\rm E}$ and $ Y_{(\ell m)ab}^{\rm
B}$ are explicitly given by
\begin{equation}
   Y_{(\ell m)ab}^{\rm E}={N_\ell\over 2} \left( \begin{array}{cc}
   W_{\ell m} & X_{\ell m} \sin\theta\\
   \noalign{\vskip6pt}
   X_{\ell m}\sin\theta & -W_{\ell m}\sin^2\theta \\
   \end{array} \right), \quad
   Y_{(\ell m)ab}^{\rm B}={N_\ell\over 2} \left( \begin{array}{cc}
   -X_{\ell m} & W_{\ell m} \sin\theta \\
   \noalign{\vskip6pt}
   W_{\ell m}\sin\theta & X_{\ell m}\sin^2\theta \\
   \end{array} \right),
\end{equation}
where
\begin{equation}
     W_{\ell m}(\hat n) \pm i X_{\ell m}(\hat n) = \sqrt{{ (l+2)!
     \over (l-2)!}}\,_{\pm2}Y_{\ell m}(\hat n),
\end{equation}
in terms of the spin-2 harmonics ${}_{\pm2}Y_{\ell m}$ used in
\citet{Seljak:1996gy}, \citet{Zaldarriaga:1996xe}, and \citet{Hu:1997hp}.
If we replace $(Q,U)$ by $(U,-Q)$, then ${\rm E}\rightarrow
-{\rm B}$ and ${\rm B}\rightarrow {\rm E}$.  This tells us
therefore, that a pure-E polarization pattern becomes a pure-B
pattern if we rotate each polarization vector by $45^\circ$, and
{\it vice versa}, as can be also inferred from the flat-sky treatment.  
Examples of E and B type polarization patterns are shown in
{\bf Figure \ref{fig:pureGC}}.  The parity properties of T, E, and B
found in the flat-sky treatment remain valid on the full sky.

\section{E and B modes from gravitational waves}  

We now return to the polarization pattern induced by a single gravitational
wave, of $+$ polarization, of wavelength $k$ propagating in the
$\hat z$ direction.  The upshot of Section \ref{sec:GWstoQU}  is
that this gravitational wave induces a polarization tensor
\citep{Kosowsky:1994cy},
\begin{equation}
     {\cal P}^{ab}_{\boldsymbol k, +}(\theta,\phi) = \frac{T_0}{4\sqrt{2}}
     \sum_\ell(2\ell+1)P_\ell(\cos\theta)\tilde\Delta_{Q \ell} \left(
     \begin{array}{cc}
     (1+\cos^2\theta)\cos2\phi & 2\cot\theta \sin2\phi \\
     2\cot\theta \sin2\phi    & -(1+\cos^2\theta)\csc^2\theta\cos2\phi
\end{array}
\right).
\label{eqn:Pab}
\end{equation}
If we expand this in tensor spherical
harmonics, the resulting coefficients are
\citep{Kamionkowski:1996ks,Zaldarriaga:1996xe},
\begin{equation}
     a^{\rm E\, \boldsymbol k, +}_{\ell m} =  \frac{\sqrt{\pi(2\ell+1)} }{4(\delta_{m,2}+\delta_{m,-2})^{-1}}  
     \left[\frac{(\ell+2)(\ell+1)
     \tilde\Delta_{Q,\ell-2}}{(2\ell-1) (2\ell+1)}+\frac{6
     \ell(\ell+1) \tilde\Delta_{Q\ell}}{(2\ell+3)(2\ell-1)} +
     \frac{\ell(\ell-1) \tilde\Delta_{Q,\ell+2}}
     {(2\ell+3)(2\ell+1)}\right],
\end{equation}
and 
\begin{equation}
     a^{\rm B\, \boldsymbol k, +}_{\ell m}= \frac{-i}{2\sqrt{2}} \sqrt{\frac{2\pi}{(2\ell+1)}}
     (\delta_{m,2}-\delta_{m,-2}) \left[(\ell+2)\tilde\Delta_{Q,{\ell-1}}+(\ell-1)
     \tilde\Delta_{Q,\ell+1} \right].
\end{equation}
We have thus shown explicitly that both the E and B components are
nonzero for a gravitational wave, confirming the heuristic arguments above.

This particular gravitational wave (in the $\hat{z}$ direction
with $+$ polarization) contributes 
\begin{equation}
     C_\ell^{{\rm BB},\, \boldsymbol k, +} = \frac{1}{2l+1} \sum_m|a^B_{\ell m}|^2
     = \frac{\pi}{2}
     \left [\frac{\ell+2}{2\ell+1}\tilde\Delta_{Q,{\ell-1}}+\frac{\ell-1}{2\ell+1}
     \tilde\Delta_{Q,\ell+1}\right]^2.
\label{eqn:EEresult}
\end{equation}
to the BB power spectrum, and similarly for $C_\ell^{\rm EE}$,
with the replacement B$\rightarrow$E in
Equation \ref{eqn:EEresult}.  Since $C_\ell^{\rm BB}$ is a
rotationally invariant quantity, any gravitational wave of this
wavenumber $k$ pointing in any direction, with either
polarization, will contribute similarly to $C_\ell^{\rm BB}$.  We
thus obtain the BB power spectrum from the stochastic
gravitational-wave background by summing all Fourier modes,
$\int d^3k/(2\pi)^3$, and over both gravitational-wave
polarization states.  The final result for $C_\ell^{\rm BB}$ is
thus,
\begin{equation}
     C_\ell^{\rm BB}=\frac{1}{2\pi}\int\, k^2\, dk
     \left[ \frac{\ell+2}{2\ell+1}\tilde\Delta_{Q,\ell-1}(k) +
     \frac{\ell-1}{2\ell+1} \tilde
     \Delta_{Q,\ell-1}(k)\right]^2,
\label{eqn:Bpowerspectrum}
\end{equation}
and analogously for $C_\ell^{\rm EE}$.
Note that the cross-correlation power spectrum vanishes,
 $C_\ell^{\rm EB}= \sum_{m=-\ell}^{m=\ell}(a^{{\rm E}*}_{\ell
 m}a^{\rm B}_{\ell m})/(2\ell+1)=0$,  as it should,
because $a_{(\ell m)}^{\rm E} \propto
(\delta_{m,2}+\delta_{m,-2})$, while $a_{(\ell m)}^{\rm B} \propto
(\delta_{m,2}-\delta_{m,-2})$ for a $+$ polarization
gravitational wave propagating in the $\hat z$ direction, and
similarly for $C_\ell^{\rm TB}$.

\begin{figure}
\includegraphics[width=\linewidth]{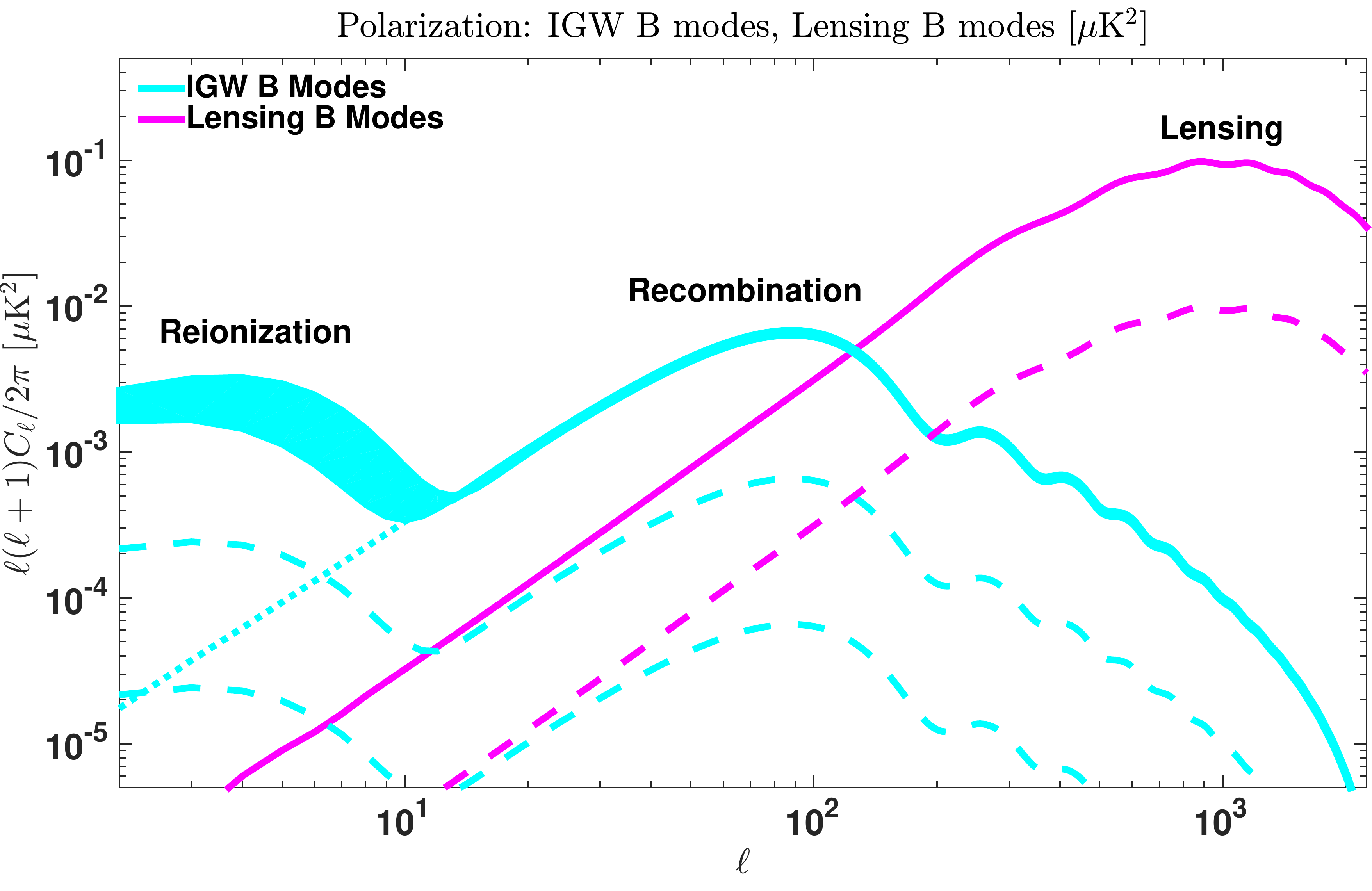}
\caption{{\it Polarization power}: Spectra are shown for primordial
     B modes with $r=\{0.1, 0.01, 0.001\}$ ({\it cyan}), lensing-induced
     B modes ({\it magenta}), as well as scalar E modes ({\it red}), for comparison. 
     The $\pm1\sigma$ uncertainty due to the current constraint on $\tau$, the 
     optical depth to reionization, is indicated for the $r=0.1$ case 
     by the ({\it cyan}) shading (the dotted line is the result with no reionization). 
     A 90\% delensed signal is also shown for comparison ({\it dashed-magenta}). 
     Plots were generated using CAMB \citep{Lewis:1999bs} with
     Planck 2015 best-fit parameters \citep{Ade:2015xua}.}
\label{fig:clbylens}
\end{figure}

{\bf Figure \ref{fig:clbylens}} shows results of numerical evaluation of
Equation \ref{eqn:Bpowerspectrum} using CAMB \citep{Lewis:1999bs}, with
the Planck 2015 cosmological parameters \citep{Ade:2015xua}.  The
``recombination peak'' in the power spectrum [multiplied by
$\ell(\ell+1)/2\pi$] at $\ell\sim100$ arises from gravitational waves
that enter the horizon around the time of CMB decoupling at
redshift $z\simeq1100$. The power drops at smaller $\ell$ because
longer-wavelength modes were superhorizon at the time
of decoupling and thus have a suppressed effect on subhorizon
physics.  The power drops at higher $\ell$ because the amplitudes
of shorter-wavelength gravitational waves, which entered the
horizon earlier, have begun redshifting away by the time of CMB
decoupling.  The ``reionization bump'' at $\ell\lesssim 10$
\citep{Ng:1994sv,Zaldarriaga:1996ke}
arises from re-scattering of the CMB by free electrons that were
reionized at redshift $z\sim8$ by ultraviolet radiation from the
first stars.  The wiggles at higher $\ell$ arise from the
difference in phases of gravitational waves at different
wavelengths at the time of CMB decoupling
\citep{Pritchard:2004qp,Flauger:2007es}.  The overall amplitude
scales with the tensor-to-scalar ratio $r$.

\paragraph{Line-of-sight and total-angular-momentum formalisms}
Numerical solution to the hierarchy of coupled integro-differential
equations required to evolve the $\Delta_{Q \ell}(k)$ forward in
time from the primordial Universe until today is computationally
intensive.  They can be formally integrated
\citep{Zaldarriaga:1995gi} to provide line-of-sight expressions
\citep{Seljak:1996is}
\begin{eqnarray}
     C_\ell^{\rm EE} &=&  \int d\ln k\, \Delta_h^2(k) \left\{
     \int\limits_0^{\eta_0-\eta_{\rm ls}}\!\!\!\!  dx g(\eta_0-x)
     \Psi(\eta_0-x) \left[- j_\ell(x)  + j_\ell''(x) + \frac{2
     j_\ell(x)}{x^2} +  \frac{4 j_\ell'(x)}{x} \right]
     \right\}^2 \nonumber \\ 
     C_\ell^{\rm BB} &=&  \int d\ln k\, \Delta_h^2(k) \left\{
     \int\limits_0^{\eta_0-\eta_{\rm ls}}\!\!\!\! dx g(\eta_0-x)
     \Psi(\eta_0-x) \left[ 2 j_\ell'(x) +\frac{4 j_\ell(x)}{x}
     \right] \right\}^2,
\label{eqn:losexpressions}
\end{eqnarray}
for the EE and BB power spectra, respectively.  Here
$x$ is a distance along the line of sight and $\eta_0$ and
$\eta_{\rm ls}$ the conformal times today and at last scatter,
respectively.  Also, $g(\eta) d\eta$ is the fraction of CMB photons that last scatter
between conformal times $\eta$ and $\eta+d\eta$.  The strange
combinations of spherical Bessel functions can be understood as
follows.  The transverse-traceless part of the three-dimensional
spatial metric---i.e., the part that describes gravitational
waves---can be decomposed into three-dimensional {\it total
angular momentum} (TAM) waves \citep{Dai:2012bc}, eigenfunctions
of wavenumber $k$ of the Laplacian and of total (orbital plus
spin) angular momentum quantum numbers $\ell$ and $m$.  There
are two such sets of TAM waves that can be labeled E and B.  The
arrangements of spherical-Bessel functions that appear in
Equations \ref{eqn:losexpressions} are then the coefficients
[Equation 94 in \citet{Dai:2012bc}] in the projection from the
three-dimensional E and B TAM waves onto the respective
two-dimensional polarization tensor spherical harmonics on the
surface of the sky.

\begin{textbox}
\subsubsection{E and B modes from density perturbations}
The same lines of reasoning that conclude that both E and B
modes are induced by gravitational waves demonstrate that
density perturbations do not produce a curl.
Consider a single Fourier mode of the density field in
the $\hat{z}$ direction.  Then the Sachs-Wolfe effect induces an
intensity variation proportional to
$(\cos^2\theta-1/3)$.  Any electron at the surface of last
scatter now sees a quadrupolar intensity variation that is
aligned with the $\hat\theta$ direction or the direction
perpendicular.  We thus find that for $\boldsymbol{k}||\hat{z}$,
$U(\hat{n})=0$, so
\begin{equation}
     {\cal P}^{ab}(\theta,\phi)= \sum_{\ell} \sin^2\theta
     \tilde\Delta^s_{Q\ell}
     P_\ell(\cos\theta)\left(\begin{array}{cc} 1 & 0\\ 0 &
     -\csc^2\theta
     \end{array} \right),
\end{equation}
where here $\tilde\Delta^a_{Q\ell}$ are the polarization moments
for a density perturbation. One finds from this polarization
pattern $a^{\rm E}_{\ell m}\neq 0,$ but $a^{\rm B}_{\ell m}= 0$.  This
happens because ${\cal P}^{ab}\epsilon^c{}_b=0$ which follows since
${\cal P}^{ab}$ is diagonal and independent of $\phi$.
Therefore, a curl component (B mode) in the CMB arises at linear
order in perturbation theory only from primordial gravitational
waves. 

\end{textbox}

\section{Lensing-induced B modes}
\label{sec:lensing}

Above we showed that density perturbations do not induce a curl
in the polarization.  However, that derivation assumed only
linear perturbations, in which each Fourier mode of the density
field is considered independently. B modes may still arise from 
the density at higher order.

The largest relevant nonlinear effect in the CMB is weak
gravitational lensing (or cosmic shear) of the primordial CMB
temperature-polarization pattern by density perturbations
between us and the CMB surface of last scatter
\citep{Zaldarriaga:1998ar,Lewis:2006fu}.  Lensing
displaces the temperature and polarization from a given
direction $\boldsymbol \theta$ at the surface of last scatter to an
adjacent position $\boldsymbol\theta + \delta\boldsymbol\theta$:
\begin{equation}
     \left(\begin{array}{c} T\\ Q\\ U\\ \end{array}
     \right)_{\rm obs.}(\boldsymbol{\theta}) =\left(\begin{array}{c}
     T\\ Q\\ U\\ \end{array}
     \right)_{\rm ls}(\boldsymbol{\theta}+\delta\boldsymbol{\theta})
     \simeq\left(\begin{array}{c} T\\ Q\\ U\\ \end{array}
     \right)_{\rm ls}(\boldsymbol{\theta}) + \delta\boldsymbol{\theta} \cdot
     \nabla\left(\begin{array}{c} T\\ Q\\ U\\ \end{array}
     \right)_{\rm ls}(\boldsymbol{\theta}),
\end{equation}
where $\delta\boldsymbol{\theta} =\nabla\Phi$ is the lensing
deflection, and $\Phi$ is a projection of the
three-dimensional gravitational potential $\Phi(\boldsymbol x)$ along
the line of sight $\hat{n}$.

The generation of B modes by lensing is most easily seen in the
flat-sky limit.  If there is no B mode at the surface of last
scatter, then (cf., Equation \ref{eqn:EBtoQUFouriercomponents})
$\tilde Q(\boldsymbol{\ell})=2 \tilde E(\boldsymbol{\ell})\cos2\varphi_{\boldsymbol \ell}$ and
$U(\boldsymbol{\ell})=-2E(\boldsymbol{\ell})\sin2\varphi_{\boldsymbol \ell}$.  Thus,
\begin{equation}
     \nabla Q(\boldsymbol\theta) = -2i\, \int\,
     \frac{d^2\boldsymbol{\ell}}{(2\pi)^2}\, 
     \tilde E(\boldsymbol{\ell})\, \cos2\varphi_{\boldsymbol \ell}\, 
     \boldsymbol{\ell}\, e^{-i\boldsymbol{\ell}\cdot\boldsymbol{\theta}},
\end{equation}
and similarly for $\nabla U(\boldsymbol\theta)$ with $\cos\rightarrow
-\sin$.  The deflection angle is likewise
\begin{equation}
     \tilde \Phi(\boldsymbol\theta)=-i\int\, \frac{d^2\boldsymbol{\ell}}
     {(2\pi)^2}\, \Phi(\boldsymbol{\ell})\,
     e^{-i\boldsymbol{\ell}\cdot\boldsymbol{\theta}}\, \boldsymbol{\ell}.
\end{equation}
Thus, the perturbation to $Q$ and $U$ induced by gravitational waves is
\begin{equation}
     \delta Q(\boldsymbol{\theta})=(\nabla Q)\cdot(\nabla
     \Phi)=\int\frac{d^2\boldsymbol{\ell}}{(2\pi)^2}
     e^{i\boldsymbol{\ell}\cdot\boldsymbol{\theta}}(\nabla Q\cdot\nabla
     \Phi)_{\boldsymbol{\ell}},
\end{equation}
where
\begin{equation}
     \delta Q(\boldsymbol{\ell})\equiv \left[(\nabla Q)\cdot(\nabla
     \Phi) \right]_{\boldsymbol{\ell}} = 2 \int \frac{d^2\boldsymbol{\ell}_1}{(2\pi)^2}
     [\boldsymbol{\ell}_1\cdot(\boldsymbol{\ell} - \boldsymbol{\ell}_1)] \tilde E(\boldsymbol{\ell}_1)
     \tilde \Phi(\boldsymbol{\ell}-\boldsymbol{\ell}_1)\cos2\varphi_{\boldsymbol{\ell}_1},
\end{equation}
\begin{equation}
     \delta U(\boldsymbol{\ell})\equiv \left[(\nabla U)\cdot(\nabla
     \Phi)\right]_{\boldsymbol{\ell}}=-2\int\frac{d^2\boldsymbol{\ell}_1}
     {(2\pi)^2}[\boldsymbol{\ell}_1\cdot(\boldsymbol{\ell}-\boldsymbol{\ell}_1)]
     \tilde E(\boldsymbol{\ell}_1)
     \tilde \Phi(\boldsymbol{\ell}-\boldsymbol{\ell}_1)\sin2\varphi_{\boldsymbol{\ell}_1}.
\end{equation}
Although the original map had (by assumption) no curl, the
lensed map does; from Equation \ref{eqn:GCFouriercomponents},
\begin{equation}
     B(\boldsymbol{\ell})=\frac{1}{2}[\sin2\varphi_{\boldsymbol{\ell}\,}\delta Q(\boldsymbol{\ell})-
     \cos2\varphi_{\boldsymbol{\ell}\,}\delta U(\boldsymbol{\ell})]
     =\int\frac{d^2\boldsymbol{l_1}}{(2\pi)^2}
     [\boldsymbol{\ell}_1\cdot(\boldsymbol{\ell}-\boldsymbol{\ell}_1)]E(\boldsymbol{\ell}_1)
     \Phi(\boldsymbol{\ell}-\boldsymbol{\ell}_1)\sin2\varphi_{\boldsymbol{\ell}_1}.
\end{equation}
If the power spectrum for the projected potential is
$C_\ell^{\Phi\Phi}$, then the B-mode power spectrum from
lensing is,
\begin{equation}
     C_\ell^{\rm BB}=\int\frac{d^2\boldsymbol{l_1}}{(2\pi)^2}
     [\boldsymbol{\ell}_1\cdot(\boldsymbol{\ell}-\boldsymbol{\ell}_1)]^2\sin^22
     \varphi_{\boldsymbol{\ell}_1} C^{\Phi\Phi}_{|\boldsymbol{\ell}-\boldsymbol{\ell}_1|}
     C_{{\ell_1}}^{\rm EE}.
\end{equation}

In {\bf Figure \ref{fig:clbylens}} we show numerical results for the
lensing B-mode power spectrum
\citep{Zaldarriaga:1998ar,Kesden:2002ku,Knox:2002pe}.  Given
the extraordinary current precision of the standard $\Lambda$CDM
parameters, the predictions for this B-mode power spectrum are
very precise, especially at the the $\ell\sim10-100$ range where the
gravitational-wave signal peaks. [Neutrino masses
\citep{Abazajian:2013oma} and/or the effects of nontrivial dark
energy may affect lensing-induced B-mode power at larger
$\ell$ \citep{Benson:2014qhw}.]  These lensing-induced B modes
have now been detected in the range $200 \lesssim \ell \lesssim
1500$ by several experiments.  Detection through
cross-correlation with tracers of the projected potential (see
below) was reported by SPTPol \citep{Hanson:2013hsb}, POLARBEAR
\citep{Ade:2013hjl}, and ACTPol \citep{vanEngelen:2014zlh}, and
then detections in autocorrelation were reported by POLARBEAR
\citep{Ade:2014afa}, BICEP2/Keck \citep{Ade:2014xna}, and SPTPol
\citep{Keisler:2015hfa}.

\paragraph{Delensing}

As is clear from {\bf Figure \ref{fig:clbylens}}, if $r$ is large
enough, then the recombination peak in the B-mode power spectrum
will stand out (given sufficiently precise experiments) from the
lensing power spectrum, and even more so the reionization bump.
There are, however, prospects to distinguish the B modes due to
lensing from those due to gravitational waves.  Measurement of
higher-order temperature-polarization correlations induced by
lensing can be used to reconstruct
the deflection angle, $\delta\boldsymbol\theta(\boldsymbol{\theta}) = \nabla
\Phi(\boldsymbol\theta)$,  as a
function of position on the sky
\citep{Zaldarriaga:1998te,Seljak:1998aq,Hu:2001kj,Kesden:2003cc},
and these may then be used to reduce the lensing-induced curl
\citep{Kesden:2002ku,Knox:2002pe,Seljak:2003pn}.  Heuristically, lensing will stretch the CMB
temperature-polarization patterns in some small region of the
sky and thus induce a local departure from statistical isotropy,
a preferred direction in the temperature-polarization
correlations over some small patch of sky.
Lensing-reconstruction algorithms then map this local departure
from statistical isotropy as a function of position on the sky.

We illustrate this by explaining how lensing reconstruction works
with a temperature map and then discuss the generalization to
polarization.  In the absence of lensing, each Fourier
mode $\tilde T(\boldsymbol \ell)$ of the temperature field is
statistically independent, $\VEV{\tilde T(\boldsymbol{\ell}) \tilde T(\boldsymbol{\ell}')} =
0$, for $\boldsymbol{\ell}\neq\boldsymbol{\ell}'$.
However, if there is lensing, an observed Fourier mode
$\tilde T(\boldsymbol{\ell})$ has contributions from all pairs of temperature and
projected-potential Fourier modes $\tilde T(\boldsymbol{\ell}_1)$ and
$\tilde \Phi(\boldsymbol{\ell}_2)$ that have
$\boldsymbol{\ell}=\boldsymbol{\ell}_1+\boldsymbol{\ell}_2$. Thus, with lensing,
\begin{equation}
     \VEV{\tilde T(\boldsymbol{\ell_1}) \tilde T(\boldsymbol{\ell_2})}=f(\boldsymbol{\ell_1},\boldsymbol{\ell_2})
     \tilde \Phi(\boldsymbol{\ell}) \:\:\:{\rm for}\:\:\: \boldsymbol{\ell_1}\neq\boldsymbol{\ell_2},
\end{equation}
in the presence of some fixed projected potential
$\Phi(\boldsymbol{\theta})$ with Fourier components
$\tilde \Phi(\boldsymbol{\ell})$. Here, $f(\boldsymbol{\ell_1},\boldsymbol{\ell_2}) =
C_\ell^{\rm TT}(\boldsymbol{\ell}\cdot \boldsymbol{\ell_1})+C^{\rm 
TT}_\ell(\boldsymbol{\ell}\cdot\boldsymbol{\ell_2})$.  Each $\boldsymbol
\ell_1$-$\boldsymbol\ell_2$ pair of observed temperature modes, with
$\boldsymbol\ell_1+ \boldsymbol \ell_2 =\ell$, then
provides an estimator $\widehat \Phi(\boldsymbol \ell) = \tilde
T(\boldsymbol \ell_1) \tilde T(\boldsymbol \ell_2)/f(\boldsymbol \ell_1,\boldsymbol \ell_2)$
for $\tilde \Phi(\boldsymbol\ell)$. The optimal estimator for
the Fourier components of projected potential is then 
obtained by adding, with
inverse-variance weighting, all the estimators
from all $\boldsymbol\ell_1$-$\boldsymbol\ell_2$ pairs with
$\boldsymbol\ell_1+\boldsymbol\ell_2 =\boldsymbol \ell$ \citep{Hu:2001kj}
\begin{equation}
     \widehat{\Phi(\boldsymbol{\ell})}=
     A(\ell) \int\frac{d^2\boldsymbol{\ell_1}}
     {(2\pi)^2} \tilde T(\boldsymbol{\ell}_1) \tilde
     T(\boldsymbol{\ell}_2)F(\boldsymbol{\ell}_1,\boldsymbol{\ell}_2),
\label{eqn:estimatorone}
\end{equation}
\begin{equation}
     F(\boldsymbol{\ell}_1,\boldsymbol{\ell}_2)\equiv\frac{f(\boldsymbol{\ell}_1,\boldsymbol{\ell}_2)}
     {2C_{\ell_1}^{\rm TT,t}C_{\ell_2}^{\rm TT,t}},\:\:\:\:\:\:\:A(L)=L^2
     \left[\int\frac{d^2\boldsymbol{\ell_1}}{(2\pi)^2}
     f(\boldsymbol{\ell}_1,\boldsymbol{\ell}_2)F(\boldsymbol{\ell}_1,\boldsymbol{\ell}_2)\right]^{-1},
\label{eqn:estimatortwo}
\end{equation}
where $C_\ell^{\rm TT,t}$ is the total observed (signal plus noise) power
spectrum. Thus, with these estimators, the projected potential
can be determined as a function of position across the sky from
the measured temperature map.  The projected-potential
measurement can then be used to ``delens'' the observed
polarization pattern; i.e., to reconstruct the polarization
pattern at the surface of last scatter from the (lensed)
temperature/polarization pattern that is observed
\citep{Kesden:2002ku,Knox:2002pe}.  

\begin{table}[h!]
\caption{Minimum variance filters for different lensing potential estimators. We define $\varphi_{12}\equiv\varphi_{\boldsymbol{\ell}_1}-\varphi_{\boldsymbol{\ell}_2}$, where $\varphi_{\boldsymbol{\ell}_1}, \varphi_{\boldsymbol{\ell}_2}$ are the angles between $\boldsymbol{\ell}_1, \boldsymbol{\ell}_2$ and the $\boldsymbol x$ axis, respectively.}
\begin{center}
\begin{tabular}{c|c|c|c}
\hline \hline
${\rm X} {\rm X}'$      & $f_{\rm XX'}({\boldsymbol{\ell}_1,\boldsymbol{\ell}_2})$ & ${\rm X} {\rm X}'$      & $f_{\rm XX'}({\boldsymbol{\ell}_1,\boldsymbol{\ell}_2})$\vsp \hline
${\rm TT}$       & $ C_{\ell_1}^{\rm TT}(\boldsymbol{\ell} \cdot \boldsymbol{\ell}_1) +
C_{\ell_2}^{\rm TT}(\boldsymbol{\ell} \cdot \boldsymbol{\ell}_2) $& ${\rm EE}$   & $[ C_{\ell_1}^{\rm EE}(\boldsymbol{\ell} \cdot \boldsymbol{\ell}_1) +
C_{\ell_2}^{\rm EE}(\boldsymbol{\ell} \cdot \boldsymbol{\ell}_2)]\cos{2\varphi_{12}}$ \vsp 
${\rm T E}$      & $ C_{\ell_1}^{\rm TE}(\boldsymbol{\ell} \cdot \boldsymbol{\ell}_1)\cos
2\varphi_{12} +  C_{\ell_2}^{\rm TE}(\boldsymbol{\ell}
\cdot \boldsymbol{\ell}_2)$ & ${\rm EB}$    & $[ C_{\ell_1}^{\rm EE}(\boldsymbol{\ell} \cdot \boldsymbol{\ell}_1) -
C_{\ell_2}^{\rm BB}(\boldsymbol{\ell} \cdot \boldsymbol{\ell}_2)]\sin
2\varphi_{12}$  \vsp 
${\rm T B}$     & $ C_{\ell_1}^{\rm TE}(\boldsymbol{\ell} \cdot \boldsymbol{\ell}_1)\sin 2
\varphi_{12}$ & ${\rm BB}$  & $[ C_{\ell_1}^{\rm BB}(\boldsymbol{\ell} \cdot \boldsymbol{\ell}_1)+
C_{\ell_2}^{\rm BB}(\boldsymbol{\ell} \cdot \boldsymbol{\ell}_2)] \cos
2\varphi_{12}$  \vsp
\hline\hline
\end{tabular}
\end{center}
\label{table:filters}
\end{table}

Similar estimators that use polarization, as well as temperature, can be
constructed analogously \citep{Hu:2001kj}.  There are then in
addition to the TT estimator described above, EE, TE, EB, TB,
and BB estimators, with coupling coefficients $f(\boldsymbol
\ell_1,\boldsymbol \ell_2)$ as given in {\bf Table \ref{table:filters}}.
The precision with which 
$\tilde\Phi(\boldsymbol{\ell})$ can be reconstructed depends on the number of
small-scale coherence patches in the temperature-polarization
map that can be used as `sources' with which the shear can be
reconstructed.  Thus, high angular resolution and high
sensitivity are required.  Since the polarization power spectrum
peaks at $\ell\sim 1000$, rather than $\ell\sim 200$, there are more
small-scale coherence patches in the polarization than in the
temperature.  Given that there are (under the null hypothesis of
no gravitational waves) no B modes in the primordial map, there
is no cosmic-variance contribution to the EB lensing estimator,
and this turns out to ultimately provide the most power in
lensing reconstruction \citep{Hu:2001kj}.  Thus, a
high-sensitivity and high-resolution polarization map is
required to optimize lensing reconstruction.  

CMB lensing has recently entered the era of detection.
The effects of lensing of the WMAP temperature map were
discovered through cross-correlation
with the NVSS radio survey \citep{Smith:2007rg} and several other
tracers of the mass distribution \citep{Hirata:2008cb} and then
through autocorrelation in ACT \citep{Das:2011ak} and SPT
\citep{vanEngelen:2012va}.  An all-sky map of the
projected  mass density was more recently constructed from
lensing of the Planck maps \citep{Ade:2013tyw}.  The effects of
lensing on the CMB polarization have now also been
detected---these are the B-mode detections discussed above.
The possibility to trace the projected potential by using
lensing of galaxies \citep{Marian:2007sr} and 21-cm maps
\citep{Sigurdson:2005cp} have also been considered
although the latter is somewhat futuristic.

\section{Foreground contributions to B modes}

The largest IGW B-mode signal allowed by current observational
limits is $O(10)\, {\rm nK}$, and a sensitivity
to a tensor-to-scalar ratio as small as $r\sim10^{-3}$ implies a
B-mode signal four orders of magnitude smaller.  Detecting a
primordial CMB signal of this amplitude is a daunting task. As
polarization is measured in bolometer experiments by taking the difference between the
temperature at two orthogonal, co-located axes (once for Q and
then at a $45^\circ$ angle for U), uncontrolled variations in
temperature along these axes can be mistaken for proper
polarization fluctuations. Unfortunately, spread out between us
and the last-scattering surface at $z\simeq 1100$, there is a long line
of obtruding foregrounds which hinder our ability to accurately
measure temperature differences at CMB frequencies.
Measurements from the ground are obscured by contributions from
man-made electromagnetic interference and atmospheric noise,
which contribute at all frequencies. In nearby outer space,
zodiacal light (emission from the interplanetary dust cloud)
generates pollution on frequencies $\gtrsim100\,{\rm
GHz}$. Further out, there are various sources of contamination
from localized objects, including inverse-Compton scattering of
CMB photons from hot electrons in intracluster gas (the
Sunyaev-Zeldovich effect
\citep{Sunyaev:1972eq,Sunyaev:1980vz,Birkinshaw:1998qp}) at
$10-300\,{\rm GHz}$,
synchrotron emission from active galactic nuclei at
$\lesssim100\,{\rm GHz}$, as well as extragalactic dust emission
(the cosmic infrared background, the integrated effect from
high-redshift galaxies) at $\gtrsim100\,{\rm GHz}$
\citep{Ade:2013zsi}.

\subsection*{Galactic foregrounds}
For large-angular-scale polarization, however, the dominant
foregrounds are Galactic in origin, mainly in the form of
diffuse synchrotron and thermal dust emissions (free-free
emission from accelerated electrons in the ionized gas and
{\it anomalous} dust emission
\citep{Kogut:1996us,Leitch:1997dx}---which is most likely
due to electric dipole radiation from small spinning dust
grains---provide additional subdominant contributions), both of
which involve the Galactic magnetic field. Together they render the
Galactic plane virtually unusable for cosmological observations,
leaving the sky at high Galactic latitudes as the focus of CMB
analysis. In order to extract a CMB signal as clean as possible,
a considerable portion of the sky is masked.  Multi-frequency measurements are
then used in order to separate the components of the radiation
in the remaining regions, relying on the fact that their
intensities differ in frequency dependence.

\paragraph{Synchrotron}
Galactic synchrotron emission is dominant at frequencies below
$100\,{\rm GHz}$, and both WMAP and Planck have observed its
polarization signature at frequencies from $30$ to $90\,{\rm
GHz}$ [up until then the only all-sky template was the Haslam
map \citep{1982A&AS...47....1H}, at a much lower frequency of $408\,{\rm
MHz}$]. These multi-frequency measurements have been used to fit
a spectral brightness temperature index of $\beta_s\sim-3$ above
$20\,{\rm GHz}$ \citep{Adam:2015tpy}. As many
of the upcoming CMB polarization experiments intend to take data
at $\lesssim90\,{\rm GHz}$, improved understanding of this foreground is
essential.

\paragraph{Dust}
Above $100\,{\rm GHz}$, thermal emission from asymmetric dust
grains in the interstellar medium, which align themselves with
the Galactic magnetic field, induces a strong polarization
signal, which depends on the composition, shape and size of the
grains \citep{Draine:2008hu,Martin:1971,AliHaimoud:2012ds,Andersson}.
Early templates for Galactic dust were based on  
smoothed maps of starlight polarization \citep{Page:2006hz}. 
While starlight polarization has been demonstrated to be a good tracer 
of dust polarization \citep{Ade:2014aaa}, this approach is limited by the sparsity 
of the data and the fact that the stars reside at different distances. 
Other templates have relied heavily on models for 
the Galactic magnetic field, and over the
years several such models have been developed [e.g.,
\citep{O'Dea:2011kx,Delabrouille:2012ye}]. 
In the absence of solid observational data, however, theoretical
templates inevitably involve considerable guesswork, and 
theoretical templates of Galactic dust-emission
foregrounds before the advent of high-frequency data from the
Planck satellite turned out to underestimate the
amplitude of dust polarization at high Galactic latitudes.

\begin{textbox}

\subsection{BICEP2}
In March 2014 the BICEP2
Collaboration reported detection of B-mode power at 150 GHz and
$\ell\sim40-100$ in excess of that expected from lensing
\citep{Ade:2014xna}.  The B-mode signal was found to have no
correlation with existing dust-polarization templates and to
have an amplitude in excess of that expected from dust.
The potentially extraordinary implications of this measurement
attracted considerable scrutiny, and arguments were made that
uncertainties in the various dust templates may have been
underestimated \citep{Mortonson:2014bja,Flauger:2014qra}.  Data
from Planck
\citep{Adam:2014bub} on polarized dust emission at high Galactic
latitude then indicated that pre-Planck dust models had
underestimated the polarization.  A subsequent joint
analysis of BICEP2/Keck and Planck data discovered a significant correlation
between the BICEP2/Keck 150 GHz polarization map and the Planck 353
GHz map, indicating that the entire BICEP2 B-mode excess could
be attributed to dust, leaving an upper limit $r\leq 0.12$ (95\% 
C.L.) to the tensor-to-scalar ratio \citep{Ade:2015tva}.

\end{textbox}

The Planck High Frequency Instrument (HFI) has recently provided
full-sky temperature and polarization maps at
frequencies ranging from $100$ to $857\,{\rm GHz}$
(corresponding to $3\,{\rm mm}$ to $350\,{\rm \mu m}$
wavelengths). Focusing on $353\,{\rm GHz}$ data at high Galactic
latitudes, the E and B angular power spectra of dust
polarization were constrained in the multipole range
$40<\ell<600$ \citep{Adam:2014bub}. The frequency
dependence was found to be consistent with a modified blackbody
emission with power-law emissivity
$\epsilon_\nu\propto\nu^{\beta_d}$ and temperature $T_d$ with
best-fit values $\beta_d = 1.59$ and $T_d = 19.6\, {\rm K}$. It was also
shown that both $C_\ell^{\rm EE}$ and $C_\ell^{\rm BB}$ spectra are
well described by power laws with exponents $\alpha_{EE,BB} =
-2.42 \pm 0.02$, almost independent of the region of sky. The
amplitudes, however, were shown to vary considerably
across the sky. While no region of sky was found to be clean
enough to enable IGW detection without foreground subtraction,
Planck did identify in each Galactic hemisphere several patches of sky 
at high Galactic latitudes with considerably lower foreground 
amplitudes, possibly useful targets for future IGW B-mode searches.

\paragraph{Dust-polarization puzzles}
While dust and lensed E modes provide a satisfactory fit to all
current measurements, several results remain
unexplained. For example, both WMAP \citep{Page:2006hz} and
Planck \citep{Adam:2014bub}
have found a systematic difference in amplitude between dust E
and B modes, roughly $C_\ell^{\rm BB}=0.5 C_\ell^{\rm EE}$, almost
independent of Galactic latitude.  This disagrees with the
expectation $C_\ell^{\rm BB}\simeq C_\ell^{\rm EE}$ 
if the polarization orientation is coherent over large
patches with small-scale modulations in amplitude
\citep{Zaldarriaga:2001st,Kamionkowski:2014wza}. It has been postulated
that this could be explained by magnetohydrodynamic turbulence
in the interstellar medium \citep{HirataFace}.

In addition, the frequency dependence of the dust polarization
fraction as observed by Planck [see {\bf Figure 13} in
\citet{Ade:2014iii}] has an opposite trend compared to the
longstanding predictions from models of silicate or carbonaceous
dust grains (see {\bf Figure 8} of \citet{Draine:2008hu}).  Models
involving magnetic nanoparticles (namely ferromagnetic or
ferrimagnetic iron grains) \citep{Draine:2012zu} may explain the
observed increase of polarization dust fraction with frequency.

It is also reasonable to wonder whether foreground polarizations
measured at different frequencies trace the same depths in the
ISM, and thus have the same polarization pattern on the sky (a
tacit assumption in such analyses).  The correlation between the
BICEP2 150 GHz maps and the Planck 353 GHz maps
\citep{Ade:2015tva} suggest that they do to some extent, but the
detailed validity of this assumption warrants further study.

\section{The Search for B modes}
\label{sec:experiment}

We now consider the prospects to achieve an experimental
sensitivity to a tensor-to-scalar ratio $r\sim 10^{-3}$ and
discuss experimental issues and strategies.

\subsection{Detectability basics}
In principle, an experiment provides the polarization Stokes
parameters $Q$ and $U$ at each point on the sky from which the
$2\ell+1$ spherical-harmonic coefficients $a_{\ell m}^{\rm BB}$
are then obtained from Equation \ref{GCmomentseasy}.  The theory
predicts that each of these $a_{\ell m}$ is drawn from a
Gaussian distribution with variance $C_\ell^{\rm BB}=
\VEV{|a_{\ell m}^{\rm BB}|^2}$.  We thus
construct an estimator $\widehat{C_\ell^{\rm
BB}}=\sum_{m=-\ell}^\ell |a_{\ell m}|^2/(2\ell+1)$, and this
estimator has a root-variance $\left(\Delta C_\ell^{\rm BB}
\right)= [2/(2\ell+1) ]^{1/2} C_\ell^{\rm BB}$ (since the
root-variance with which we can measure the variance $\sigma^2$
of a Gaussian distribution from $N$ measurements is $(2/N)^{1/2}
\sigma^2$).

In practice, things are complicated by detector noise in the
measurement, imperfectly subtracted foreground contributions to
the CMB polarization, and the fact that measurements may be
available only over a fraction $f_{\rm sky}$ of the sky.  If we
are interested to detect IGW-induced B modes, there is also
contamination from lensing-induced B modes.  The estimator for
$C_\ell^{\rm igw}$ (for notational economy, we drop the BB 
superscript in the remainder of this section) is then obtained by subtracting from the
measured $C_\ell$ the expected contributions $C_\ell^{\rm lens}$,
$C_\ell^{\rm fg}$, and $C_\ell^{\rm n}$ from lensing,
imperfectly subtracted foregrounds, and detector noise,
respectively.  The root-variance with which we can measure
$C_\ell^{\rm igw}$ then becomes
$(\Delta C_\ell) = [2/(2\ell+1) f_{\rm sky}]^{1/2}\left(C_\ell^{\rm igw} +
C_\ell^{\rm lens} + C_\ell^{\rm fg} + C_\ell^n \right)$.
The increase by $f_{\rm sky}^{-1/2}$ in the root variance arises
from the decrease in sky coverage. 

To evaluate the detectability of IGW B modes, we parametrize the
IGW-induced B-mode power spectrum $C_\ell^{\rm igw} = 10\,r\,
C_\ell^{\rm igw}(r=0.1)$ in terms of the power 
spectrum $C_\ell^{\rm igw}(r=0.1)$ for a tensor-to-scalar
ratio $r=0.1$, and an amplitude $r$.  Each
measured multipole moment $C_\ell^{\rm obs}$ then provides an
estimator $\widehat{r}^\ell = 0.1\,(C_\ell^{\rm obs} -
C_\ell^{\rm lens}-C_\ell^{\rm fg}-C_\ell^{\rm n})/C_\ell^{\rm
IGW}(r=0.1)$ to the amplitude $r$.  The error to this
estimator, under the null hypothesis $r=0$, is
$(\sigma_r^\ell)^2 =0.01\, [2/(2\ell+1)] (C_\ell^{\rm
lens}+C_\ell^{\rm n} +C_\ell^{\rm fg})^2$.  These estimators can
then be added with inverse-variance weighting to obtain the
minimum-variance estimator for $r$.  This estimator will
have a root-variance,
\begin{equation}
    \sigma_r \simeq \frac{0.1}{f_{\rm sky}}\,\left( \sum_{\ell=
    \ell_{\rm min}}^{\ell_{\rm max}}
    \frac{(2\ell+1)}{2}
    \left[ \frac{C_\ell^{\rm IGW}(r=0.1)}{C_\ell^{\rm
    fg}+C_\ell^{\rm n}+  C_\ell^{\rm lens}}
    \right]^2  \right)^{-1/2},
\label{eqn:idealsensitivity}
\end{equation}
for an experiment that covers a fraction $f_{\rm sky}$ of the sky and measures multipole moments from
a minimum $\ell_{\rm min}$ to a maximum $\ell_{\rm
max}$. 

\begin{figure}
\includegraphics[width=0.8\linewidth]{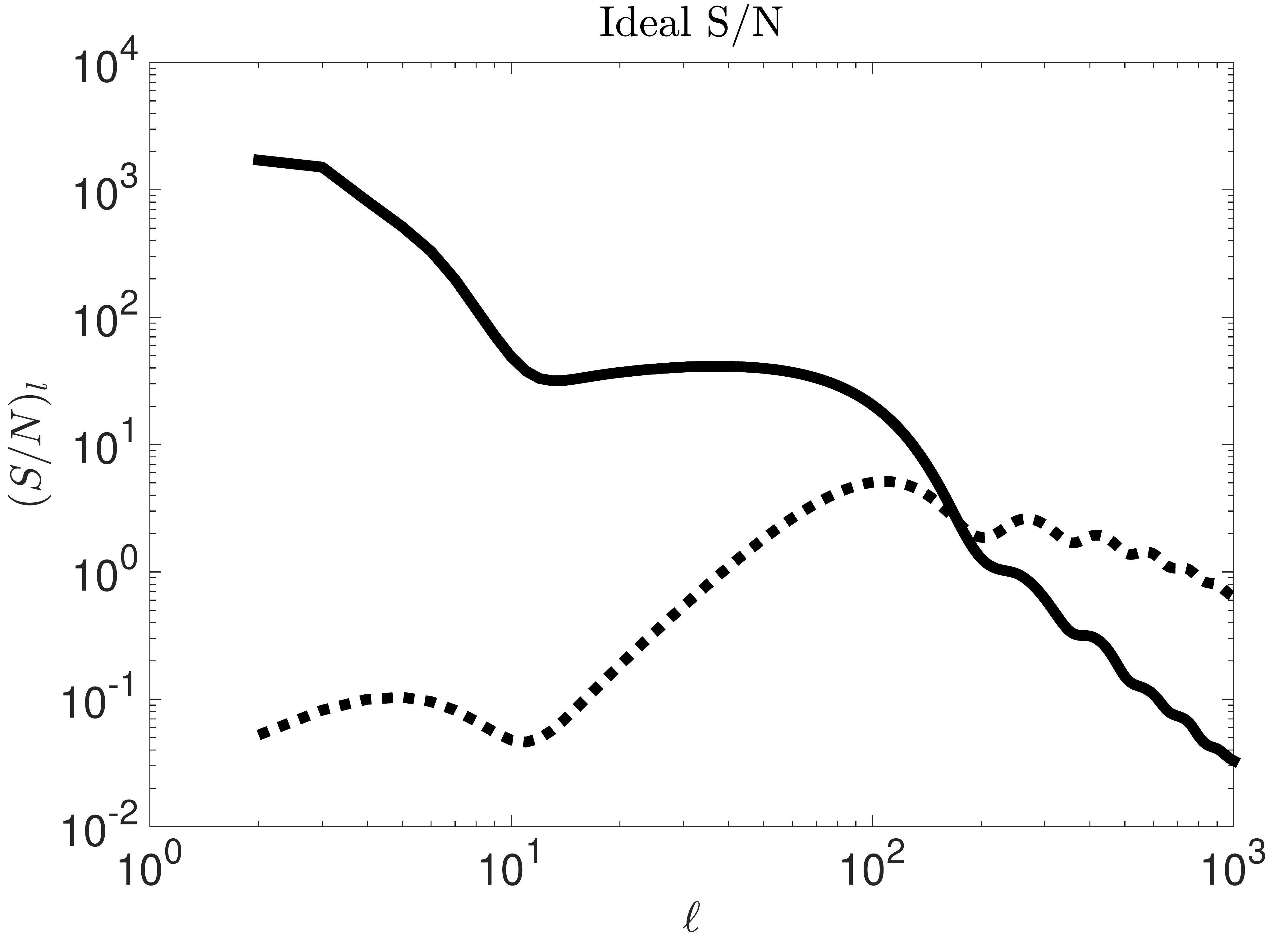}
\caption{The summand from
    Equation \protect\ref{eqn:idealsensitivity}, for
    the detectability of IGW B modes, with (i) $C_\ell^{\rm
    fg}=0$---i.e., in the absence of foregrounds but with
    detector noise and lensing-induced B modes
    included---and (ii) $C_\ell^{\rm  n}=C_\ell^{\rm
    lens}=0$---i.e., in the absence of
    lensing-induced B modes and detector noise.
    This figure shows the contributions of various multipole
    moments to the signal-to-noise assuming the hypothetical
    situations of (i) no foregrounds, or (ii) no detector
    noise nor lensing.}
\label{fig:summands}
\end{figure}

\subsection{The effects of different contaminants}

\subsubsection{Lensing-induced B modes}
Before including the effects of foregrounds and detector
noise, let us first consider the signal-to-noise available in
the hypothetical situation of a no-noise and
foreground-free measurement; i.e., $C_\ell^{\rm n}=C_\ell^{\rm
fg}=0$.  In this case there would be a cosmic-variance limit to
the measurement of $r$ from lensing-induced B modes, and
{\bf Figure \ref{fig:summands}} plots the summand in
Equation \ref{eqn:idealsensitivity}  as a function of $\ell$.
As shown there, the vast majority ($\gtrsim99\%$) of the
signal-to-noise on the sky is in the reionization bump, at $\ell
\lesssim 10$ \citep{Kamionkowski:1997av}; only $\sim1\%$ is at 
$\ell\gtrsim10$.  If we were to restrict measurements to 
$\ell\gtrsim10$, the vast majority of the signal-to-noise is at
$\ell\lesssim150$ (the recombination bump in $C_\ell^{\rm IGW}$).

Evaluating the sum in Equation \ref{eqn:idealsensitivity}, we can 
calculate $\sigma_r$, the smallest detectable 
(at $1\sigma$) tensor-to-scalar ratio $r$, for experiments with access to
different multipole ranges. For $2\leq\ell\leq10$, we get $\sigma_r\simeq 
4\times 10^{-5}\,(\tau/0.078)^{-2}$, while for $10\lesssim\ell\lesssim150$
we get $\sigma_r\simeq 3\times 10^{-4}\,f_{\rm sky}^{-1/2}$ 
\citep{Lewis:2001hp,Knox:2002pe,Kesden:2002ku}.  
Here we have included the $\propto \tau^{-2}$ scaling with the reionization
optical depth \citep{Ng:1994sv,Zaldarriaga:1996ke} for the
low-$\ell$ measurement and the $f_{\rm sky}^{-1/2}$
scaling for the high $\ell$.  Although (with no foregrounds) the
vast majority of the signal-to-noise is at $\ell \lesssim 10$, 
an experiment that maps only $\sim\!1\%$ of
the sky (and thus uses only $\ell\gtrsim10$) could still have a
$\gtrsim3\sigma$ sensitivity to a tensor-to-scalar ratio as small
as $r\sim 0.01$, even with no delensing.  As we will see
below, delensing by a factor of 10 is conceivable with
forthcoming experiments in which case $r\sim 0.001$ could be
detected at $\gtrsim3\sigma$, even with $f_{\rm sky}\sim1\%$.

\subsubsection{Detector noise}
We now include the effects of detector noise, still ignoring
foregrounds.  The power
spectrum induced by detector noise is $C_\ell^{\rm n} =
4\pi\,f_{\rm sky}  {\rm NET}_{\rm array}^2 / t_{\rm obs}$ (for
multipole moments $\ell \lesssim \theta_{\rm fwhm}^{-1}$
accessible with a beam width $\theta_{\rm fwhm}$), where
the noise equivalent temperature (NET) of the array is defined
as ${\rm NET}_{\rm array}=s/\sqrt{N_{\rm det}}$ in terms of the
noise-equivalent temperature $s$ of each detector and the number
$N_{\rm det}$ of detectors in the array.  Here $t_{\rm obs}$ is
the integration time on this fraction $f_{\rm sky}$ of the sky.
We now note that the lensing-induced power spectrum
is also well approximated by a constant $C_\ell^{\rm lens} \simeq
1.8\times 10^{-6}~\mu$K$^2$ over the range $\ell \lesssim 150$ of
interest here. (Powers are also sometimes quoted as a
sensitivity $\left( C_\ell \right)^{1/2}$, wherein the
lensing power is 4.6~$\mu$K-arcmin.)  The distribution of
signal-to-noise with $\ell$ is, with detector noise, thus
again exactly as shown in {\bf Figure \ref{fig:summands}}.

The $f_{\rm sky}$ scaling of $C_\ell^{\rm n}$ for an experiment
that limits its observations to a fraction $f_{\rm sky}$ of the
sky implies that a lower $C_\ell^{\rm n}$ can be achieved by
integrating deeply on a smaller patch of sky.  This
has important implications \citep{Jaffe:2000yt,Keating:BICEP}, as
discussed below, for the choice of the fraction of the sky
surveyed, especially for many of the suborbital experiments that
cannot access $\ell \lesssim 10$.  

\subsubsection{Foregrounds}
As discussed above, the principal foregrounds---synchrotron and
dust emission from the Milky Way---can be disentangled using
measurements of the polarization at multiple frequencies.  There
will, however, always be some residual foreground contribution
to any realistic CMB polarization map.  Here we use the scalings,
$C_\ell^{\rm fg} \propto \ell^{-2.42}$ determined empirically
for WMAP and Planck for $\ell\gtrsim 40$, and we assume that
this scaling extends down to $\ell\lesssim 10$ (although the validity of
this assumption is unknown). We set the foreground amplitude 
to the best-fit values measured by Planck at $353\,{\rm GHz}$ 
on scales $40<\ell<370$ over the BICEP2 field, extrapolated down
to $150\,{\rm GHz}$ (see \citep{Kovetz:2015pia} for details on the extrapolation). 
The contributions to the
signal-to-noise with which a nonzero value of $r$ can be
distinguished from the null hypothesis $r=0$, assuming no
detector noise nor lensing, are then shown in
{\bf Figure \ref{fig:summands}}.  They are distributed more evenly in
$\ell$ than the summands assuming no foregrounds, with a
significant peak at $\ell\sim100$.

\subsection{Experimental strategies} Above we saw roughly
speaking the effects of different contaminants on the IGW
detectability.  These considerations are ingredients in the
development of experimental strategies to detect IGWs.
These ingredients must then be amalgamated with a number of
logistical/experimental/hardware issues---e.g., detector
technologies and sensitivities; atmospheric frequency windows;
the availability of telescopes; funding constraints; etc.---in
the development of experiments.  Here we discuss some of the
issues, illustrate some of the tradeoffs, and summarize some of
the strategies currently being considered.

\begin{figure}
\includegraphics[width=\linewidth]{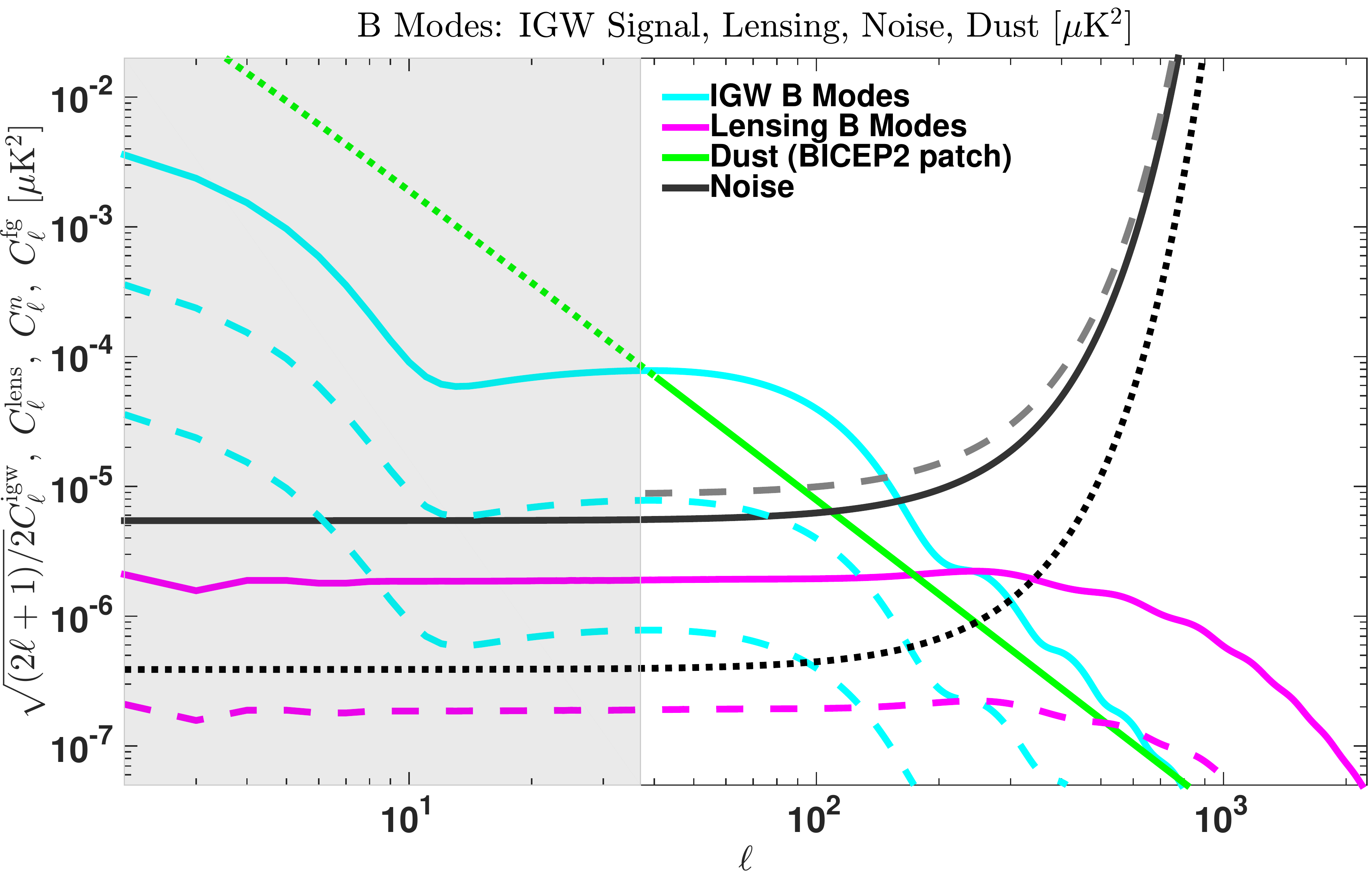}
\caption{{\it Balance of Power}: Spectra with
    $\sqrt{(2\ell+1)/2}$ scaling are shown for primordial
    B modes with $r=\{0.1, 0.01, 0.001\}$ ({\it cyan}), lensing
    of primordial E modes ({\it magenta}) and a fiducial dust foreground
    ({\it green}).  The amplitude and shape of this fiducial foreground
    were set to the best-fit values measured by Planck at $353\,{\rm GHz}$ 
    on scales $40<\ell<370$ over the BICEP2 field, extrapolated down
    to $150\,{\rm GHz}$. Whether this extends
    to larger scales is unknown. Noise spectra are plotted (without
    the $\sqrt{(2\ell+1)/2}$ scaling) for the CLASS ({\it solid
    black}), CMB-S4 ({\it dotted-black}) and BICEP2 ({\it dashed-gray}) 
    experiments. Scales inaccessible to BICEP2 due to its limited 
    sky coverage are shaded ({\it light gray}).}
\label{fig:PowSpec}
\end{figure}

We begin by plotting in Figure \ref{fig:PowSpec} in
a unified but somewhat unconventional way the contributions of
the various ingredients to the signal-to-noise. The IGW B-mode
signal is plotted as $\sqrt{(2\ell+1)/2}C^{\rm igw}_\ell$ for three
fiducial values for the tensor-to-scalar ratio, $r=\{0.1, 0.01,
0.001\}$.  The prefactor $\sqrt{(2\ell+1)/2}$ is chosen since
$(C_\ell^{\rm igw})^2$ is multiplied by the square of this
factor in the expression, Equation \ref{eqn:idealsensitivity}, for
the signal to noise.  The foreground, noise, and lensing B modes
signals are then plotted with {\it no} scaling $\ell$ prefactor.  In
this way, the contributions of foregrounds, noise, and lensing
to the signal-to-noise with which a nonzero value of $r$ can be
distinguished from the null hypothesis $r=0$ are represented
viscerally.

For the foregrounds, the amplitude of the plotted
signal is taken to match the best-fit value measured by Planck
over the BICEP2 field, as explained above.  The foregrounds in other regions of sky may
be smaller or larger.  The foregrounds can be reduced, though,
relative to what we have shown, with multi-frequency measurements, 
as we discuss below. For illustration, we also show noise power spectra 
for the CLASS (predicted) and BICEP2 instruments, as well as the planned
CMB-S4 sensitivity (see below), plotted as $C^n_\ell$
(i.\ e., without the $\sqrt{(2\ell+1)/2}$ scaling) on the scales
accessible given their beam size. The parameters sets
$\{t_{\rm obs}, {\rm NET}_{\rm array}, f_{\rm sky}\}$ used for the noise estimates for BICEP2, 
CLASS and CMB-S4 (assuming $10^5$ detectors) are $\{590\,{\rm days}, 18.75\,{\rm \mu K\sqrt{sec}}, 0.01\}$, 
$\{3\,{\rm yrs}, 7\,{\rm \mu K\sqrt{sec}}, 0.7\}$ and $\{2\,{\rm yrs}, 1.5\,{\rm \mu K\sqrt{sec}}, 0.75\}$, respectively.

\begin{figure}
\centering
 \begin{minipage}[b]{0.48\textwidth}
   \includegraphics[width=\textwidth]{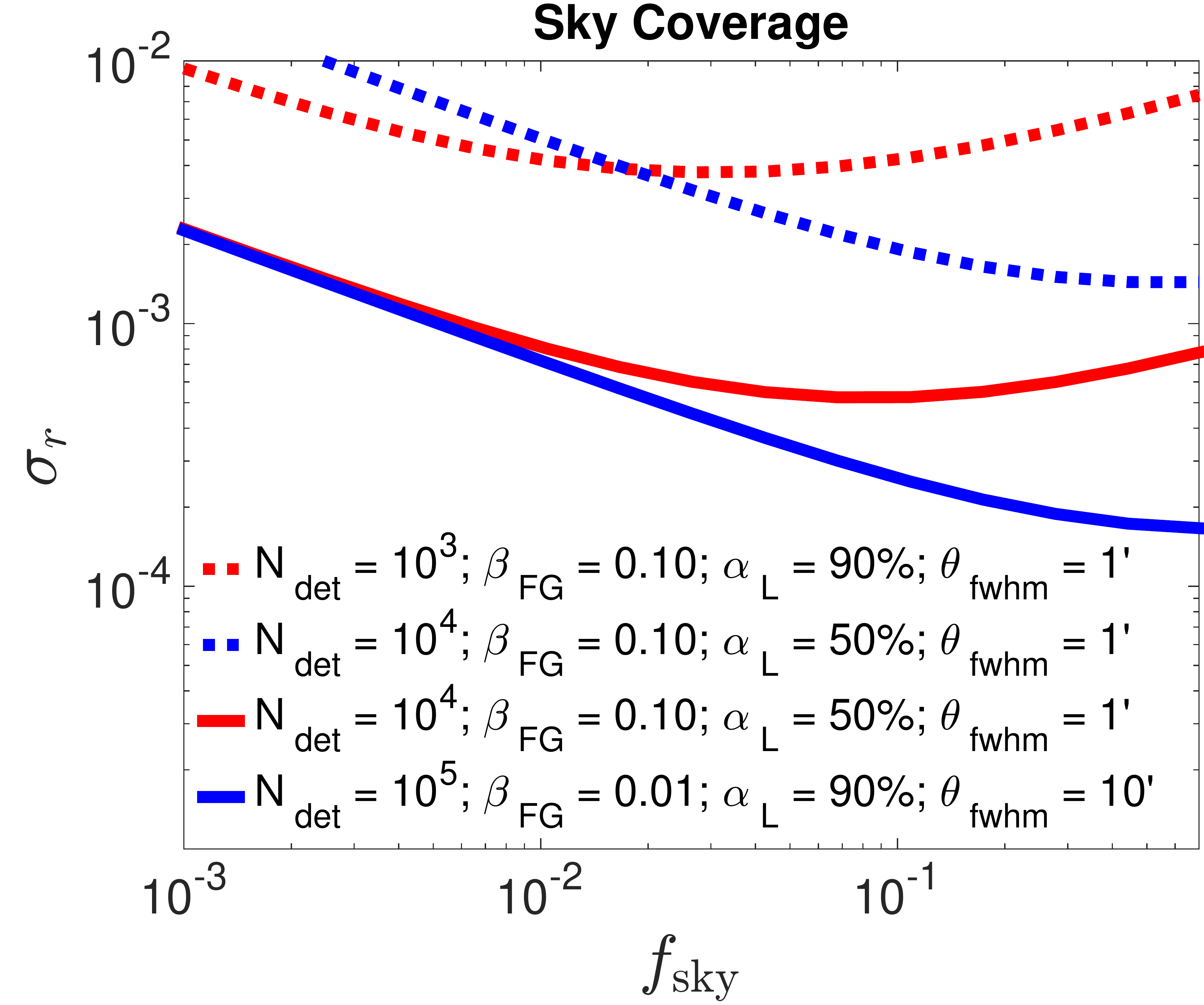}
 \end{minipage}
 \begin{minipage}[b]{0.48\textwidth}
   \includegraphics[width=\textwidth]{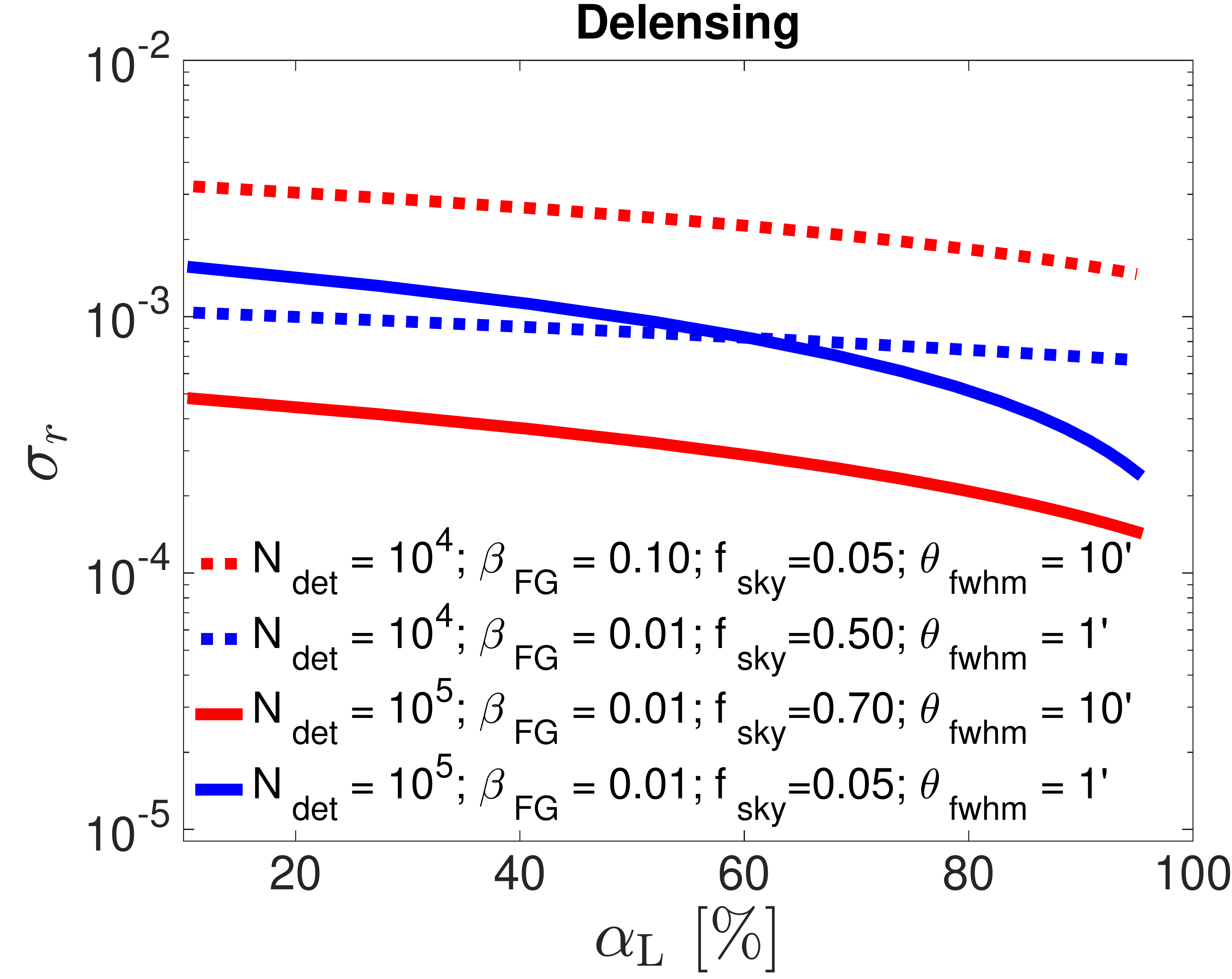}
 \end{minipage}
 \hfill
 \begin{minipage}[b]{0.48\textwidth}
   \includegraphics[width=\textwidth]{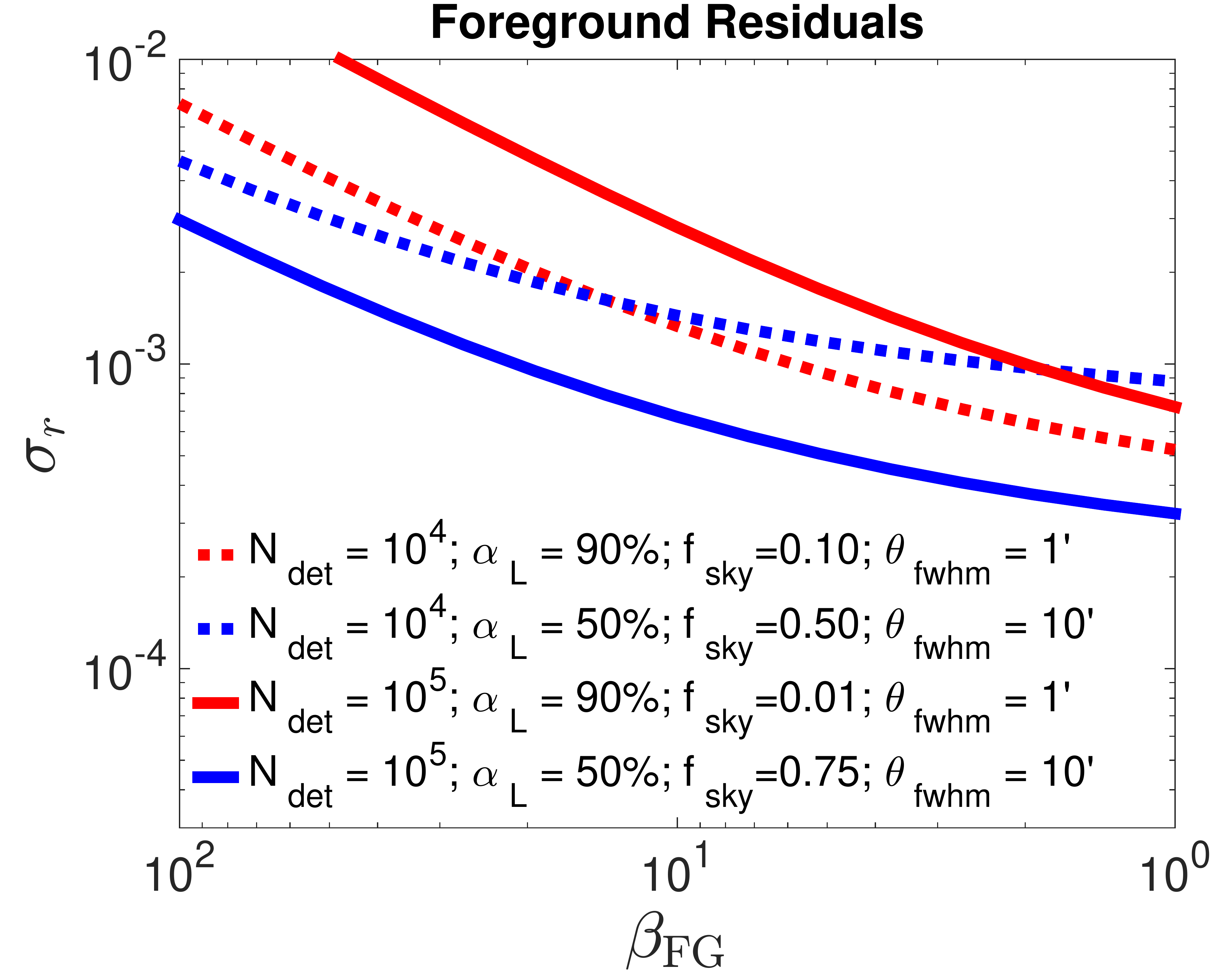}
 \end{minipage}
 \begin{minipage}[b]{0.48\textwidth}
   \includegraphics[width=\textwidth]{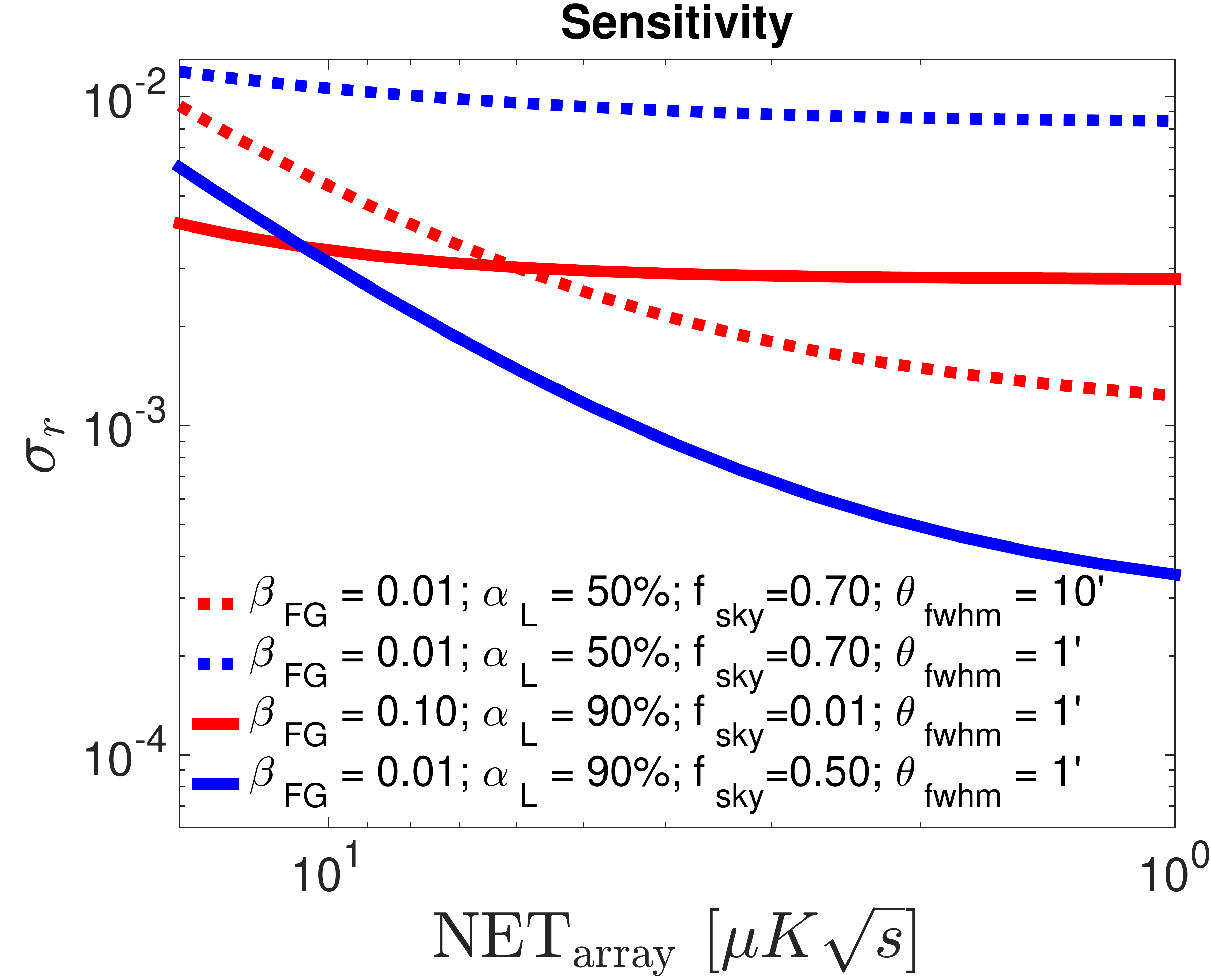}
 \end{minipage}
\caption{{\it Smallest detectable tensor-to-scalar ratio $r$}: we show
    the dependence on the sky coverage $f_{\rm sky}$, delensing
    efficiency (where $1-\alpha_L$ quantifies the lensing
    residual), foregrounds (where $\beta_{\rm fg}$ denotes the
    fractional amount of foreground residuals from the fiducial value
    we have assumed here, see {\bf Figure \ref{fig:PowSpec}}) and detector
    sensitivity (through ${\rm NET}_{\rm array}$). We show how
    the smallest detectable tensor-to-scalar $r$ (at $1\sigma$)
    depends on each parameter, while all other parameters are
    held fixed according to different scenarios, which are
    chosen to cover the range of interesting cases. We adopt the 
    instrumental sensitivity considered in \citet{Wu:2014hta},
    i.\ e. a NET per detector of $s=350\,\mu{\rm K}\sqrt{\rm sec}$,
    assume a total of $2$ years of observation, and use the
    fiducial amplitude of dust polarization from
    {\bf Figure \ref{fig:PowSpec}}.}
\label{fig:upperbound}
\end{figure}

We next calculate $\sigma_r$ 
as a function of various experimental parameters.  We evaluate
Equation \ref{eqn:idealsensitivity} under the null hypothesis
($C_\ell^{\rm igw}=0$), replacing $C_\ell^{\rm lens} \to
(1-\alpha_L) C_\ell^{\rm lens}$ and $C_\ell^{\rm fg} \to
\beta_{\rm fg} C_\ell^{\rm fg}$, where $1-\alpha_L$ parametrizes the residual lensing
contribution after delensing, and $\beta_{\rm fg}$
parametrizes the residual foregrounds after multifrequency component
separation. While imperfect subtraction can lead to a residual bias in 
$r$ \citep{Katayama:2011eh, Remazeilles:2015hpa}, note that we consider here only the
contribution to the variance. The detector noise is directly
proportional to ${\rm NET}^2_{\rm array}$, while the dependence
of $C_\ell^{\rm n}$ on $f_{\rm sky}$ introduces a tradeoff
between the detector noise and the other contributions. 
The effects of these parameters on  
$\sigma_r$ are shown in {\bf Figure \ref{fig:upperbound}}.

\subsubsection{Sky coverage}  
\paragraph{Small-sky strategy (no lensing)}  If access to the
reionization peak at $\ell\lesssim10$. is
unavailable, and in the ideal case of no lensing, the ideal
B-mode observation is achieved by prolonged integration on a
$\sim 25$-square-degree patch of sky \citep{Jaffe:2000yt}. This
is demonstrated in Figure \ref{fig:upperbound}, where there is a
sweet spot in the two cases where the detector noise is
dominant compared to the other contributions.  The sensitivity
weakens on larger patches because of the increase in
$C_\ell^{\rm n}$ with $f_{\rm sky}$.  The sensitivity then
weakens on smaller patches because the decrease in the signal
for measurements restricted to $\ell \gtrsim100$ that then miss
much of the recombination peak.

\paragraph{Lensing and the small-sky strategy} 
If an experiment has noise $C_\ell^{\rm n} \gtrsim
(1-\alpha_L) C_\ell^{\rm lens}$, then the measurement will be
detector-noise dominated, and it will be cosmic-variance
(lensing) limited otherwise.  
In the former case, the
sensitivity of the measurement to $r$ may be improved by
decreasing the sky coverage of the survey; the $f_{\rm sky}$
reduction in $C_\ell^{\rm n}$ overtakes the $f_{\rm sky}^{-1/2}$
statistical increase in the prefactor of
Equation \ref{eqn:idealsensitivity} \citep{Jaffe:2000yt}.  Still, as $f_{\rm sky}$ is
reduced, $\ell_{\rm min}$ is increased, and if $f_{\rm sky}$ is
reduced too much, then the reduction in the signal-to-noise (the
area under the curve in {\bf Figure \ref{fig:summands}})
outweighs the reduction in
$C_\ell^{\rm n}$.  

Once the detector-noise contribution to the power spectrum
is reduced to $C_\ell^{\rm n}\lesssim 2\times
10^{-6}\,\mu$K$^2\simeq C_\ell^{\rm lens}$, this small-sky
strategy must be revisited \citep{Kesden:2002ku,Verde:2005ff}.  In Figure
\ref{fig:upperbound}, it is evident, for example, that when
the detector noise is high (i.\ e., comparable to or higher
than the lensing contribution), there is nothing to gain by
delensing.  However, as we begin to access smaller
tensor-to-scalar ratios, delensing becomes increasingly
important.  Further improvements in sensitivity
to $r$  must then come from increased sky coverage, to deal with the
lensing-induced cosmic variance, or from delensing (discussed
more below), to directly reduce the lensing contribution. In
practice, the small-sky strategy has already been pursued on
larger, $\sim400$-square-degree, patches to deal with systematic
effects and foregrounds, for redundancy, and to avoid the
lensing contribution, which rises (relative to the IGW
contribution) rapidly with $\ell$.   Given the rapid
improvements in ${\rm NET}_{\rm array}$, the lensing issue will
become increasingly important.

\paragraph{The large-sky strategy} If an experiment can map the
polarization over most of the sky,
then it can access the huge amount of information in the
reionization peak, $\ell\lesssim 10$.  Lensing will not be an
issue for the foreseeable future for such an experiment, but
such a strategy will require effective isolation of
foregrounds.  See, e.g., \citet{Watts:2015eqa} for a discussion
of the large-sky strategy.

Regardless of the strategy (small-sky or large-sky), any
experiment will never be able to use data from the entire sky.
Techniques have therefore been developed to apply the E/B
decomposition on a cut sky \citep{Lewis:2001hp,Hu:2002vu,
Smith:2005gi,Pearson:2014qna}.

\subsubsection{Beam size}  Since $C_\ell^{\rm igw}$ decays
exponentially at $\ell \gtrsim 100$, high angular resolution is
not strictly needed in order to detect IGW B modes.  If the CMB
map is to be used for delensing, however, it will require very
high angular resolution, as discussed below.

\subsubsection{Detector noise}  The dependence on the
sensitivity of the instrument is more straightforward. Figure
\ref{fig:upperbound} shows how improved sensitivity can lead
to better sensitivity to $r$ given the choice of other parameters.
As discussed above, the sensitivity of the small-sky strategy is
limited only by detector noise as long as $C_\ell^{\rm n}
\gtrsim C_\ell^{\rm lens}$, and further improvements to the
sensitivity, in the absence of delensing, then drive the survey
to larger sky fractions.  Given, however, that $C_\ell^{\rm n}
\propto f_{\rm sky}^{-1/2}$, the detector noise must
continue to improve in order for the detector-noise power
$C_\ell^{\rm n}$ to continue to remain smaller than $C_\ell^{\rm
lens}$ as $f_{\rm sky}$ is increased.

\subsubsection{Frequency coverage}  The dependence on the frequency
coverage is harder to quantify and depends on the desired
sensitivity to $r$; on the relative contributions of dust and
synchrotron emission (which may depend on the region(s) of sky
covered); atmospheric windows (for terrestrial observations);
the technologies and sensitivities available at
the different frequencies; and the availability of reliable
external templates for the foreground polarization. It is clear,
though, that measurements in more frequencies over the same
patch will enable better foreground removal (as discussed above,
the main foreground contributions, due to synchrotron and dust
emission, quickly dominate as one pulls away in frequency to
either side of the CMB observability peak at $\sim100\,{\rm
GHz}$). In Figure \ref{fig:upperbound}, we examine the effect of
lower foreground residuals under several scenarios for
the remaining experimental parameters. 

\subsubsection{Delensing}  We have swept a huge amount of dirt
under the rug through the introduction of the delensing
parameter $\alpha_L$, as delensing will be an ambitious,
sophisticated, and challenging endeavor.  One possibility is
that delensing may be done with external data sets that can be
used to map the lensing potential.  For example,
measurements of the cosmic infrared background (CIB) currently
provide as good a lensing template as anything else
\citep{Sherwin:2015baa}, and forthcoming galaxy surveys, like
LSST, may reduce the lensing B modes by a factor $\sim2$ \citep{Marian:2007sr}.
However, the  most likely source for delensing at the level
required to access $r\sim0.001$  will be small-angular-scale
fluctuations in the CMB.  

Although it may ultimately be done by an experiment that also
measures the $\ell \lesssim 150$ IGW B modes, a delensing
measurement requires angular resolution far better (up to
$\ell\sim2000$) than that required for IGW B modes.  A measurement with
resolution required to reach multipole moment $\ell$ at
wavelength $\lambda$ requires a dish of size $D \sim \lambda
\ell \sim 4\,{\rm meter}\,(\ell/2000) (\nu/150\,{\rm
GHz})^{-1}$, and so the telescope-diameter requirements for
delensing are roughly ten times those for the IGW B modes.
Delensing is also optimized with high-angular-resolution maps of
the polarization, as well as temperature.  The precise level of
delensing depends on a variety of experimental parameters, as
discussed, for example, in \citet{Smith:2010gu} and
\citet{Simard:2014aqa}.  However, to
illustrate, we note that the lensing-induced B-mode power may be
reduced by a factor $\sim5$ with a polarization map with a beam
size $\sim5$~arcmin and noise level 1~$\mu$K.  The SPT-3G
project expects to be able to delens by a factor $\sim4$ by 2019
\citep{Benson:2014qhw}.

\subsection{Current/forthcoming experiments}  We now provide a
brief listing of some of the experiments underway, in
development, or being discussed.  There are several that focus a
single telescope on a chosen patch of sky to
target the recombination peak and perhaps, if the
detector-noise level warrants, delens with
higher-resolution data from the same experiment or from
external datasets.  Current and future experiments belonging to
this class include: ABS \citep{ABS}, ACTPol
\citep{Naess:2014wtr} and its successor AdvACT, the BICEP/KECK series
\citep{Ade:2014gua,Ade:2015fwj}, POLARBEAR and the future Simons
Array \citep{Arnold:2014}, and SPTPol
\citep{Hanson:2013hsb} and its successor SPT-3G
\citep{Benson:2014qhw}. A similar strategy is employed by the QUBIC 
interferometer \citep{QUBIC}.  There are then a smaller number of 
sub-orbital projects that
employ a wide--sky-coverage telescope, or aggregate several
ground-based telescopes, and pursue the reionization scales
$\ell\lesssim10$. Typically, ground-based telescopes can reach
higher resolution compared to satellites, which may enable a
more efficient delensing process. CLASS
\citep{Essinger-Hileman:2014pja}, with $\gtrsim70\%$
sky-coverage, is one experiment of this type, while under the
CMB-S4 plan, several ground-based telescopes such as those above
are planned to collaborate in generating a combined, nearly
full-sky, map \citep{Wu:2014hta,Abazajian:2013vfg}.  Balloon-borne CMB
experiments have less adaptivity and observing time than
ground-based telescopes, but they experience less atmospheric
interference and can access higher frequencies which may enable
more efficient component separation.  EBEX
\citep{ReichbornKjennerud:2010ja}, LSPE \citep{LSPE}, 
PIPER \citep{Lazear:2014bga}, and SPIDER \citep{Crill:2008rd},  
are balloon experiments, the latter two of which
target larger areas of the sky than most of the small-sky
ground-based missions.  There are then discussions of
a satellite to take full-sky polarization data and fully capture
the reionization peak. Proposals for future missions include
COrE \citep{Bouchet:2011ck}, CMBPOL \citep{Baumann:2008aj}, 
EPIC \citep{Bock:2009xw}, PIXIE \citep{Kogut:2011}, PRISM \citep{Andre:2013nfa} and LiteBIRD
\citep{Matsumura:2013aja}.  A satellite experiment could also map the
full sky with sufficiently high resolution to delens the entire
sky and thus use the IGW information also in the recombination
peak.  Such a mission would, however, require a far larger
mirror and thus be more costly.  It would, however, also enable
a broad range of interesting high-$\ell$ science apart from IGW
B modes.
More details about possible future experimental endeavors can be
found in \citet{Abazajian:2013vfg} and
\citet{Baumann:2008aj}. See also \citet{Creminelli:2015oda}. 

\subsection{Mitigating dust}
As discussed above, Planck has identified a handful of relatively clean
$\sim400$-square-degree patches of sky accessible to
observatories in the southern hemisphere.  Given Planck's noise
limitations, however, it is still
unclear whether any of these patches is far cleaner, and thus a
better B-mode target, than the others, and if so which one.  One
possibility is a brief initial high-frequency integration
\citep{Kovetz:2015pia}, either from the ground (at 220 GHz) or
353 GHz (from a balloon) to identify the cleanest such patch
before beginning a deep B-mode integration at lower (CMB)
frequencies.  Such a strategy can conceivably improve the
ultimate sensitivity to $r$ by a factor of 2--3 over a blind
selection of one of the cleanest patches.  It would also provide
high-signal-to-noise dust-polarization templates on all six of
these regions.  There are also adaptive-survey strategies
\citep{Kovetz:2013tla} that can be employed to seek
low-dust-amplitude regions while simultaneously performing a
B-mode integration.

\begin{figure*}[h!]
\centering
\includegraphics[width=0.95\linewidth]{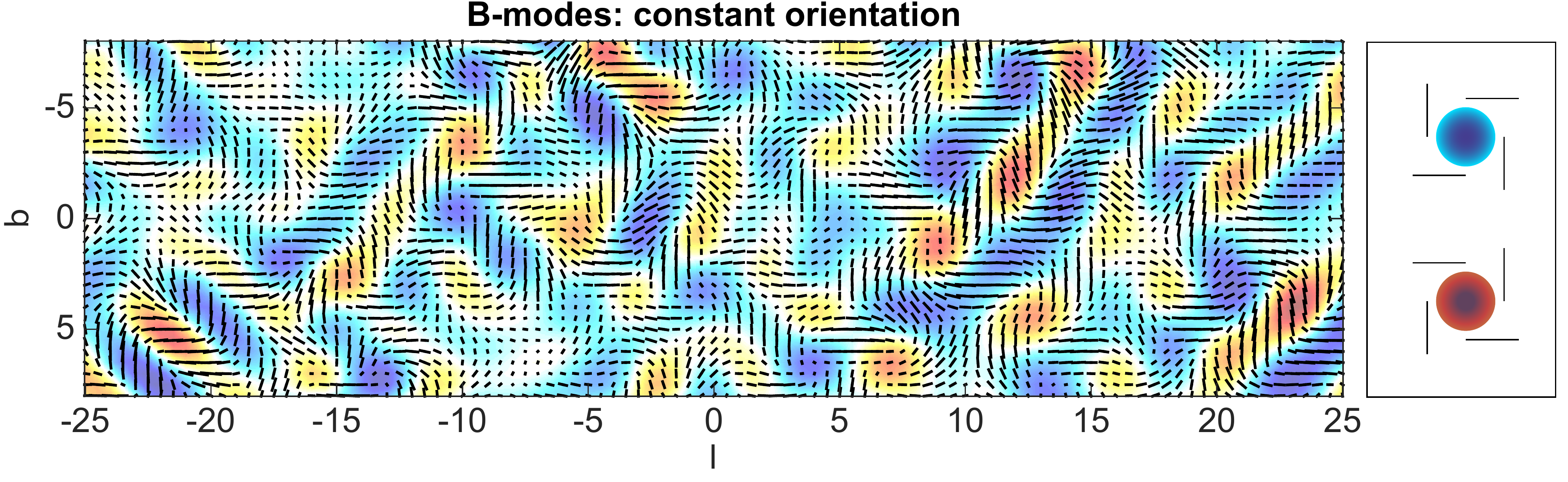}
\caption{A B-mode
     map calculated from a randomly generated, statistically
     isotropic $Q$ map and a $U=0$ map, to simulate a constant
     polarization orientation.  Here, there is a local hexadecapolar departure from
     statistical isotropy, dominated by Fourier modes oriented
     at $45^\circ$ with respect to the polarization orientation.
     More generally, the local departures from statistical
     isotropy due to a slowly-varying orientation angle can be
     captured with appropriate statistical estimators
     \protect\citep{Kamionkowski:2014wza}. (Compare with bottom of {\bf Figure \ref{fig:pureGC}}.)}
\label{fig:BmodeOrientation}
\end{figure*}

There are additional cross-checks that can be employed in
the event that a nominal IGW B-mode signal is identified even
after multifrequency component separation.
While the gravitational-wave signal is expected to be Gaussian,
the B modes from dust contamination are should be highly
non-Gaussian (as are the lensing-induced B modes; in fact it is
their characteristic non-Gaussianity that allows them to be
delensed).  If, for example, the orientation of the
dust-induced polarization is relatively coherent on large
patches of the sky, which may be expected given the large-scale
coherence of Galactic magnetic fields, then the resulting
B modes will have a locally hexadecapolar departures from
statistical isotropy, composed primarily of Fourier modes
aligned primarily in directions $45^\circ$ with respect to
the polarization orientation
\citep{Zaldarriaga:2001st,Kamionkowski:2014wza}.  Statistical
estimators to seek this type of departure from statistical
isotropy are then easily constructed
\citep{Kamionkowski:2014wza} in analogy with
lensing-reconstruction estimators. To illustrate, we  
show in {\bf Figure \ref{fig:BmodeOrientation}} the hexadecapolar
symmetry which results from having a constant orientation angle
over the observed sky patch.

\begin{textbox}

\subsubsection{Chiral gravitational waves}

As discussed above, TB and EB cross-correlations may arise if the physics 
that gives rise to CMB temperature/polarization fluctuations is parity breaking.
Chiral gravitational waves---a gravitational-wave background 
with an asymmetry between the density of right- and
left-circularly polarized gravitational waves---provide a mechanism to 
induce such parity-violating correlations \citep{Lue:1998mq,Contaldi:2008yz}.  
Chiral GWs arise if there is a Chern-Simons 
modification to gravity \citep{Jackiw:2003pm,Alexander:2009tp}
during inflation \citep{Lue:1998mq}, a parity-breaking
gravitational action during inflation \citep{Contaldi:2008yz}, or
if inflation involved Horava-Lifshitz gravity
\citep{Takahashi:2009wc}.  Chiral gravitational waves also
arise in models of inflation with a background gauge field
\citep{Maleknejad:2012fw,Adshead:2013nka}, and an analogous
mechanism could also work in the late Universe
\citep{Bielefeld:2014nza}.  The chirality of the
GW background may also be connected to the
cosmic baryon asymmetry
\citep{Alexander:2004us,Alexander:2004wk,Alexander:2004xd}.
Since IGWs induce B modes only at
multipole moments $\ell\lesssim100$, the cosmic-variance limit
to the sensitivity of any measurement to the chirality of the
IGWs is significant, and the prospects
to detect chiral gravitational waves are reasonable, only if the
tensor-to-scalar ratio $r$ is relatively large and if the
chirality is significant \citep{Gluscevic:2010vv,Ferte:2014gja}. 

\end{textbox}

\section{Other paths to inflationary gravitational waves}

There are other possibilities to detect inflationary gravitational waves.
Although these are perhaps a bit further down the road than B modes, they may
help characterize the gravitational-wave background, in case of
detection, by complementing the CMB measurement, which probes
gravitational waves with $\sim10^{-17}$~Hz frequencies, with
measurements at far higher frequencies
\citep{Smith:2005mm,Chongchitnan:2006pe,Smith:2008pf}.  The
idea to seek the inflationary background with gravitational-wave
detectors was considered in
\citet{Liddle:1993zj}, \citet{BarKana:1994bu},
\citet{Turner:1996ck}, \citet{Smith:2005mm},
\citet{Caldwell:1998aa}, and \citet{Smith:2008pf},
and has motivated mission concept studies for space-based
gravitational-wave observatories like the Big-Bang Observer
\citep{Phinneyetal,Crowder:2005nr} and DECIGO \citep{Seto:2001qf}.

There is also the
possibility to seek IGWs via their
effects on the large-scale galaxy distribution.  The
gravitational-wave background may give rise to local quadrupolar
departures from statistical isotropy in primordial perturbations
\citep{Maldacena:2002vr,Seery:2008ax,Giddings:2010nc,Giddings:2011ze,Jeong:2012df,Jeong:2012nu,Schmidt:2012ne,Bramante:2013moa}.
It may also gravitationally lens the galaxy distribution
\citep{Dodelson:2003bv,Masui:2010cz,Schmidt:2012nw,Schmidt:2013gwa}, the CMB
\citep{Cooray:2005hm,Li:2006si,Book:2011na,Dodelson:2010qu,Dai:2012bc,Dai:2013ikl}, or the 21-cm background
\citep{Pen:2003yv,Book:2011dz}, affect the intrinsic alignments of 
elliptical galaxies
\citep{Schmidt:2013gwa,Chisari:2014xia,Schmidt:2015xka} via the
tidal-alignment model \citep{Catelan:2000vm}, or have consequences
for precision astrometry of quasars \citep{Book:2010pf}.  These
effects arise in SFSR inflation but are very small, at best.
They may, however, be larger in solid inflation
\citep{Akhshik:2014bla,Dimastrogiovanni:2014ina}, non-attractor
inflation \citep{Dimastrogiovanni:2014ina}, quasi-single-field
inflation \citep{Dimastrogiovanni:2015pla}, and
globally-anisotropic models \citep{Emami:2015uva}.  Strictly
speaking, the precise distinction between early-time and
late-time effects of gravitational waves on the galaxy
distribution depends on the gauge choice, an issue explored and
clarified in recent work
\citep{Dai:2013kra,Pajer:2013ana,Dai:2015rda}.

A constraint to the primordial gravitational-wave amplitude at
frequencies $\nu \gtrsim 10^{-11}$~Hz higher than those accessible
with the CMB or large-scale structure can be obtained from
big-bang nucleosynthesis (BBN) \citep{Allen:1996vm}---such
gravitational waves would act as an additional relativistic
degree of freedom.  This bound can then be extended to
$\nu\gtrsim 10^{-15}$~Hz and improved with measurements of
small-scale CMB fluctuations, which now improve upon the BBN
bound to the number of relativistic degrees of freedom
\citep{Smith:2006nka,Sendra:2012wh,Pagano:2015hma}.  Still,
these upper bounds are probably too weak to be constraining for
SFSR inflation, although they may be of interest for models that
predict a blue ($n_t>0$) gravitational-wave spectrum \citep{Brandenberger:2014faa}.

\section{Conclusions}
We have described the quest for the B modes, the curl component
of the CMB polarization, that arise from inflationary
gravitational waves.  Until recently, the lack of existing
constraints to inflationary models allowed an almost arbitrarily
small value for the tensor-to-scalar ratio $r$ that parametrizes
the strength of the B-mode signal.  The plot has thickened in
recent years, however, with measurements that show with
increasing confidence that the scalar spectral index $n_s$
departs from unity.  Although model dependencies prevent an
absolutely conclusive statement, single-field slow-roll models
of inflation generally predict, with current constraints to
$n_s$, values of $r$ within striking distance of experimental
capabilities on a 10-year timescale. 

The challenge now will be to make these measurements, and a
massive global effort is now underway.  We summarized in Section
\ref{sec:experiment} the issues that face experimentalists and
the prospects for their resolution. There may also be room for
new ideas (e.g., to cross-correlate the reionization-bump B
modes with galaxy surveys \citep{Alizadeh:2012vy}), to
facilitate the pursuit of inflationary gravitational waves.

Of course, if $r$ is not too much smaller than the current upper
bound $r\lesssim 0.09$ (from the combination of temperature and 
polarization constraints), then a detection may be just around the
corner. If so, the obvious next step will be to characterize the
gravitational-wave background through measurements of the
spectral index $n_t$ \citep{Boyle:2014kba,Huang:2015gca}, tests
of the Gaussianity of these B modes, and the chirality
of the GW background, and perhaps with complementary
measurements of the gravitational-wave background at much
smaller wavelengths.

The prospects for a fairly definitive test of the prevailing
single-field slow-roll models of inflation have motivated
considerable efforts in the pursuit of precise measurements of
CMB polarization.  The detection of a signal, if/when it occurs,
will provide an entirely new window back to $10^{-38}$ seconds
after the cosmic singularity; provide evidence, albeit indirect,
for interesting new physics at the GUT scale; constitute a
detection of gravitational waves; and
moreover, provide the first empirical information about the
quantum behavior of the spacetime metric.  All this may occur on
a 10-year timescale, so pay attention!

\section*{DISCLOSURE STATEMENT}
The authors have nothing to disclose.

\section*{ACKNOWLEDGMENTS}
It is a pleasure to thank Daniel Baumann, Tom Crawford, Brian Keating and Eiichiro Komatsu 
for helpful suggestions and comments on an earlier version of the manuscript.
MK acknowledges the hospitality of the Aspen Center for Physics,
supported by NSF Grant No.\ 1066293.  This work was supported at
JHU by NSF Grant No.\ 0244990, NASA NNX15AB18G, the John
Templeton Foundation, and the Simons Foundation. 

%
%

\end{document}